\documentclass{article}
\usepackage[preprint]{neurips_2025}

\usepackage[utf8]{inputenc} 
\usepackage[T1]{fontenc}   
\usepackage[hidelinks]{hyperref}  %
\usepackage{url}            
\usepackage{booktabs}       
\usepackage{amsfonts}     
\usepackage{nicefrac}
\usepackage{microtype}
\usepackage{xcolor}       
\usepackage{xspace}
\usepackage{fancyvrb} %
\usepackage{fvextra}
\usepackage{float}

\usepackage{url}
\usepackage{booktabs}
\usepackage{xspace}
\usepackage{enumitem}
\usepackage[flushleft]{threeparttable}
\usepackage{makecell}
\usepackage{multirow}
\usepackage{xcolor, colortbl}
\usepackage{subcaption}
\usepackage{tikz}
\usepackage{wrapfig}
\usepackage{textcomp}
\usepackage{scalerel}  
\usepackage{tipa}
\usepackage{pifont}
\usepackage{multicol}
\usepackage[most]{tcolorbox}
\usepackage{listings}
\usepackage{caption}
\usepackage[normalem]{ulem}
\usepackage[noabbrev]{cleveref}

\newcommand{\tech}{\textsc{Live-SWE-agent}\xspace}
\renewcommand{\cite}{\citep}

\newcommand{\chatrepair}{ChatRepair\xspace}
\newcommand{\agentless}{Agentless\xspace}
\newcommand{\sweagent}{SWE-agent\xspace}
\newcommand{\minisweagent}{mini-SWE-agent\xspace}
\newcommand{\livesweagent}{\textsc{Live-SWE-agent}\xspace}
\newcommand{\hgm}{HGM\xspace}
\newcommand{\sica}{SICA\xspace}
\newcommand{\dgm}{DGM\xspace}
\newcommand{\openhands}{OpenHands\xspace}
\newcommand{\acr}{AutoCodeRover\xspace}
\newcommand{\Comment}[1]{}

\newcommand{\sbv}{SWE-bench~Verified\xspace}
\newcommand{\swebench}{SWE-bench\xspace}
\newcommand{\swebenchpro}{SWE-Bench Pro\xspace}
\newcommand{\swebenchmultilingual}{SWE-bench Multilingual\xspace}

\newcommand{\gptfivemini}{GPT-5-Mini\xspace}
\newcommand{\gptfive}{GPT-5\xspace}
\newcommand{\claudesonnetfourfive}{Claude 4.5 Sonnet\xspace}

\newcommand{\gptfivenano}{GPT-5-Nano\xspace}
\newcommand{\claudesonnetthreeseven}{Claude 3.7 Sonnet\xspace}
\newcommand{\claudesonnetfour}{Claude 4 Sonnet\xspace}
\newcommand{\geminithreepro}{Gemini 3 Pro\xspace}

\newcommand{\llm}{LLM\xspace}

\newcommand{\parabf}[1]{\vspace{.03in}\noindent \textbf{#1}}
\newcommand{\CodeIn}[1]{{\small \texttt{#1}}}

\definecolor{yucky}{HTML}{a64d79}
\newcommand*\circled[1]{\scalebox{0.8}{\tikz[baseline=(char.base)]{
\node[anchor=text, shape=circle,fill=yucky, inner sep=0pt, minimum size=1.2em] (char) {\footnotesize \textcolor{white}{#1}};}}}

\newcommand{\openai}{\scalerel*{\includegraphics{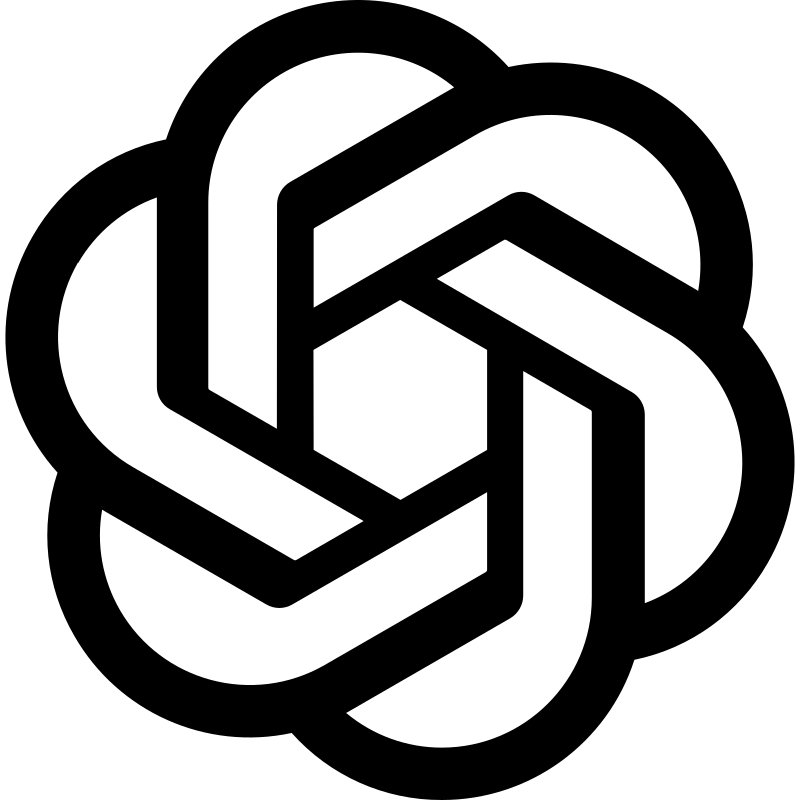}}{\textrm{C}}\xspace}
\newcommand{\anthropic}{\scalebox{1}{\scalerel*{\includegraphics{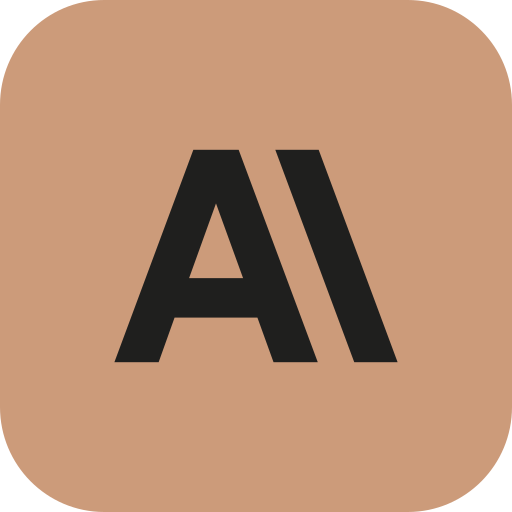}}{\textrm{C}}}\xspace}
\newcommand{\gemini}{\scalebox{1}{\scalerel*{\includegraphics{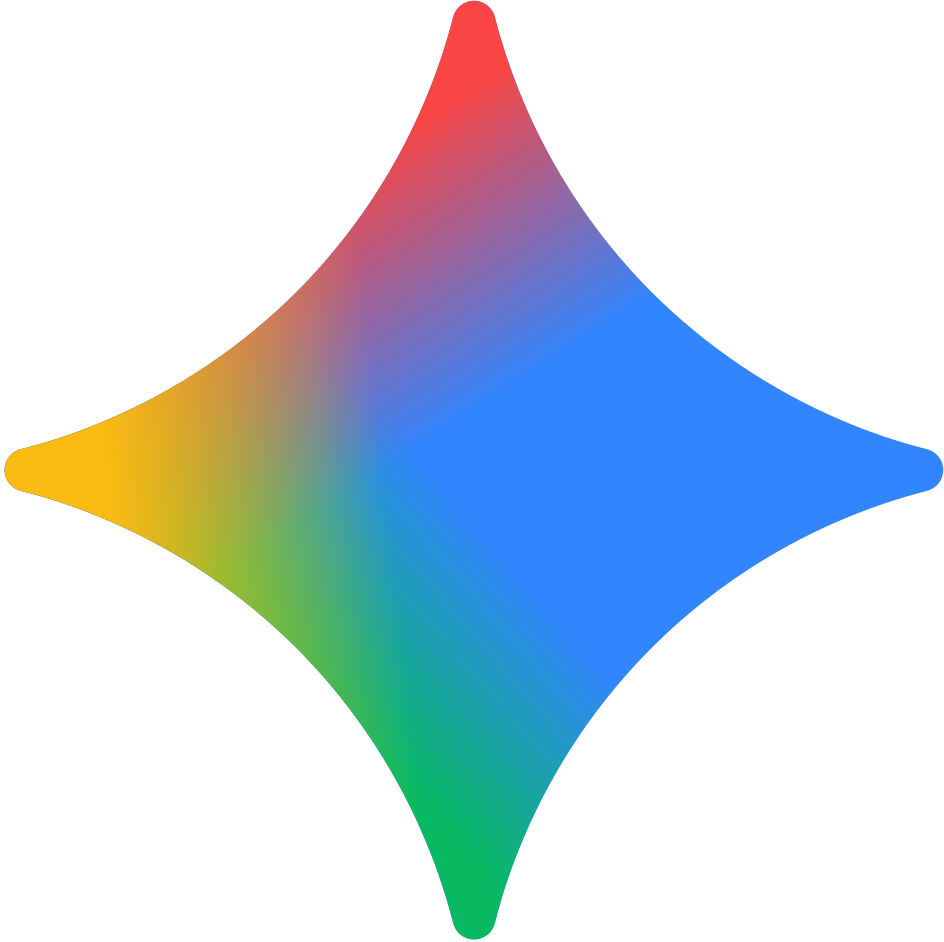}}{\textrm{C}}}\xspace}

\newcommand{\Github}{\raisebox{-1pt}{\includegraphics[height=0.8em]{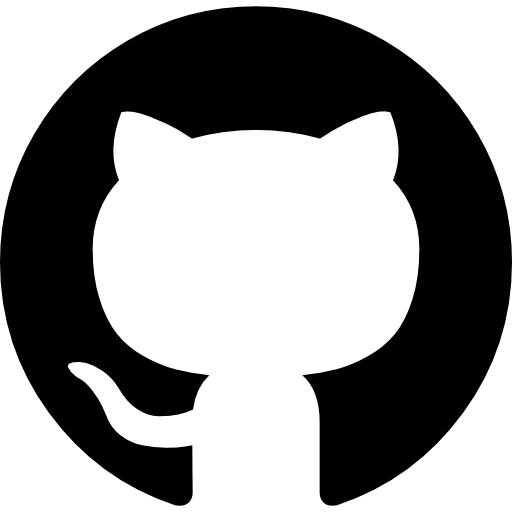}}\xspace}

\newcommand{\uiuc}[1]{{#1\textsuperscript{\includegraphics[scale=0.004]{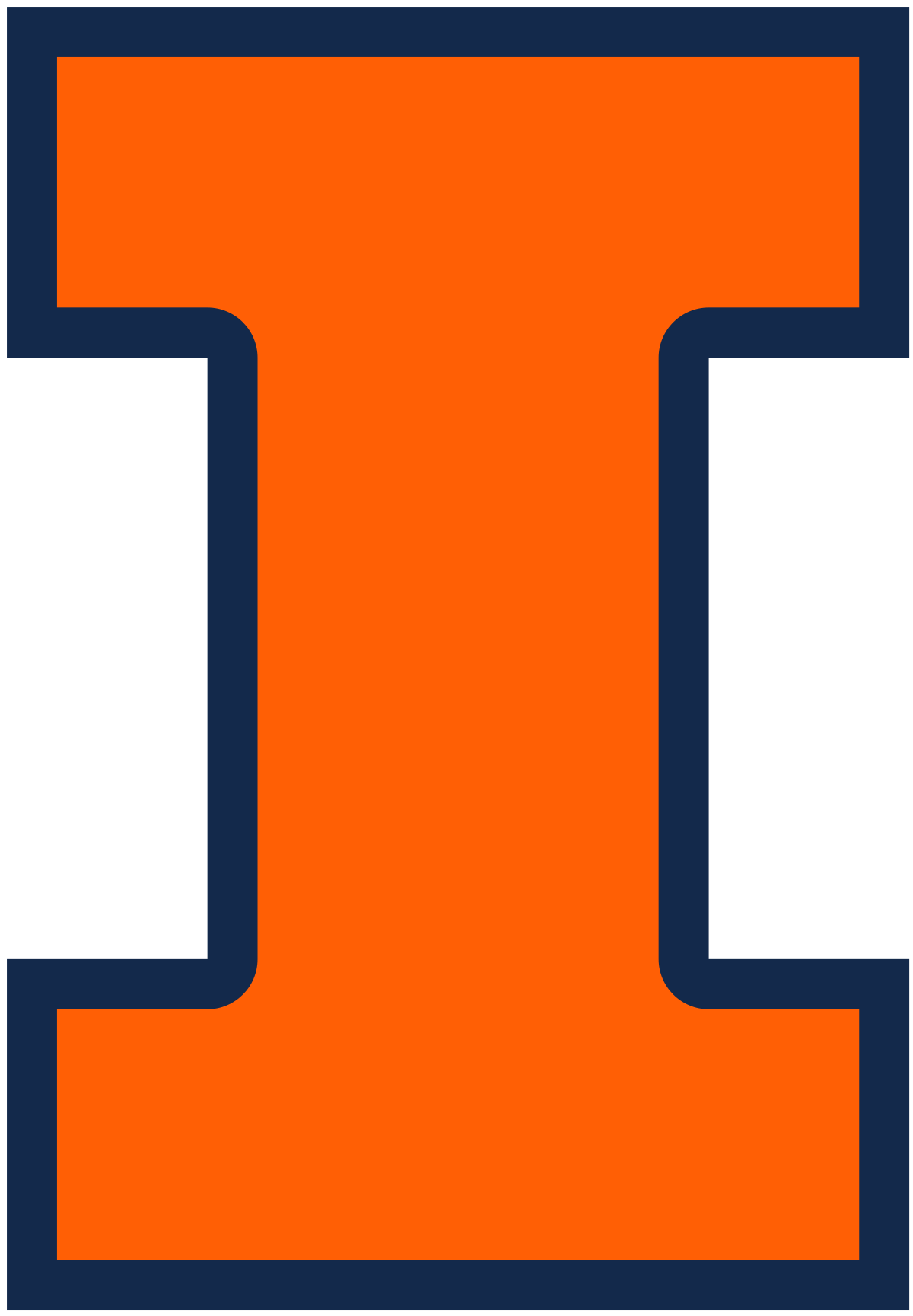}}}}

\newtcolorbox{promptbox}[1]{
  enhanced,
  breakable,
  boxrule = 1.5pt,
  fontupper = \tiny,
  fonttitle = \bf\color{white},
  arc = 5pt,
  rounded corners,
  colframe=blue!50!green,
  colback=blue!5!white,
  coltitle=white,
  title = #1,
  left=4pt %
}

\title{\livesweagent: Can Software Engineering Agents Self-Evolve on the Fly?}

\author{Chunqiu Steven Xia\;
  Zhe Wang\;
  Yan Yang$^\dagger$\;
  Yuxiang Wei\;
  Lingming Zhang
  \\[\medskipamount]
  \uiuc{University of Illinois Urbana-Champaign}
  \\[\medskipamount]
  \small\texttt{\{chunqiu2, zhe36, ywei40, lingming\}@illinois.edu} $^\dagger$\small\texttt{yanyang826@outlook.com}
}

\begin{document}

\maketitle

\makeatletter{\renewcommand*{\@makefnmark}{}\footnotetext{$^\dagger$Work done as a research intern at the University of Illinois Urbana-Champaign.}}

\frenchspacing

\begin{abstract}
Large Language Models (LLMs) are reshaping almost all industries, including software engineering. In recent years, a number of LLM agents have been proposed to solve real-world software problems. Such software agents are typically equipped with a suite of coding tools and can autonomously decide the next actions to form complete trajectories to solve end-to-end software tasks. 
While promising, they typically require dedicated design and may still be suboptimal, since it can be extremely challenging and costly to exhaust the entire agent scaffold design space. 
Recognizing that software agents are inherently software themselves that can be further refined/modified, researchers have proposed a number of self-improving software agents recently, including the Darwin-Gödel Machine (DGM). Meanwhile, such self-improving agents require costly offline training on specific benchmarks and may not generalize well across different LLMs or benchmarks. 
In this paper, we propose \livesweagent, the first \emph{live} software agent that can autonomously and continuously evolve itself \emph{on-the-fly} during runtime when solving real-world software problems. More specifically, \livesweagent starts with the most basic agent scaffold with only access to \texttt{bash} tools (e.g., \minisweagent), and autonomously evolves its own scaffold implementation while solving real-world software problems. 
Our evaluation on the widely studied \swebench Verified benchmark shows that \livesweagent can achieve an impressive solve rate of 77.4\% without test-time scaling, outperforming all existing software agents, including the best proprietary solution. Moreover, \livesweagent outperforms state-of-the-art manually crafted software agents on the recent \swebenchpro benchmark, achieving the best-known solve rate of 45.8\%. More details at: \Github{}~\small{\color{purple}\url{https://github.com/OpenAutoCoder/live-swe-agent}}

\end{abstract}

\vspace*{-11pt}
\begin{figure}[h!]
    \centering
    \includegraphics[width=1\columnwidth, height=5cm]{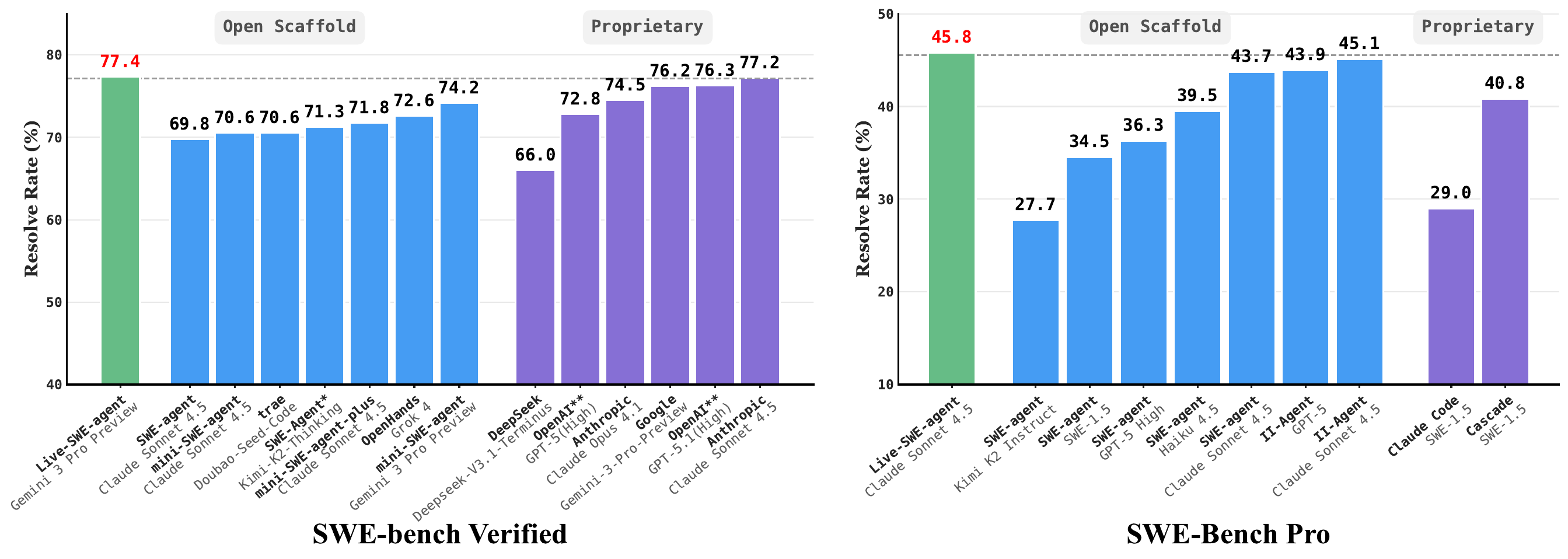}\vspace{-5pt}
    {\scriptsize\slshape {\textsuperscript{*}Customized SWE-agent ~\citep{kimik2thinking}} \quad\textsuperscript{**}JSON-based \texttt{apply\_patch} tool plus dynamic thinking time~\citep{gpt51}}\vspace{-5pt}
    \caption{\swebench Verified and \swebenchpro results (single attempt w/o test-time scaling)
    }
    \label{fig:result}
\end{figure}

\section{Introduction}
Large Language Models (\llm{s}) have rapidly progressed from simple code autocompletion~\cite{austin2021program,chen2021codex,li2022competition, xia2022less} to interactive agents capable of navigating repositories, running tests, and submitting patches end‑to‑end~\cite{yang2024sweagent,wang2024openhands,zhang2024autocoderover,xia2024agentless,carbonneaux2025cwm, liu2024large}.
Early conversational repair systems leveraged environment feedback to iteratively refine candidate fixes~\cite{xia2024automated}, while subsequent agentic frameworks (such as \sweagent~\cite{yang2024sweagent} and \openhands~\cite{wang2024openhands}) augmented LLMs with terminals, editors, and search, enabling multi‑step tool use on complex repositories. At the same time, complementary ``more agentless'' pipelines (e.g., Moatless~\cite{moatless} and \agentless~\cite{xia2024agentless}) argue that much of the perceived complexity in agent scaffold design can be replaced by dedicated prompting and workflow design.

Despite this progress, most existing agents have fixed designs and are limited in their static action space, where the scaffold implementation (including tools) is preset even when a task could benefit from more customization. In addition, manually designing an optimal software agent scaffold is extremely challenging and costly due to the infinite design space.
As a result, more recently, the community has begun to explore self-evolving software agents~\cite{zhang2025darwin,robeyns2025sica,wang2025huxley}, which iteratively modify their own scaffold implementations and empirically validate each change using offline evaluation signals from coding benchmarks.
However, these approaches add significant additional cost.
For example, a single run of \dgm on \swebench{} {costs} around \$22,000 according to the original paper~\cite{zhang2025darwin}.
Furthermore, they rely heavily on offline evolution, where improvements are typically learned on certain benchmarks and then baked into a static agent.
In this way, the learned agents may become specialized to the given benchmarks and underlying LLMs, and may not generalize well beyond that.

To bridge this gap,
we introduce \livesweagent, the first
\emph{live}, runtime self-evolving software engineering agent that expands and
revises its own capabilities \emph{on the fly} while working on a real-world issue.
Our key insight is that software agents are themselves software systems, and modern LLM-based software agents already possess the intrinsic capability to extend or modify their own implementation at runtime.
{While our idea of enabling on-the-fly self-evolution applies to all parts of the agent scaffold implementation, as a first step, we focus primarily on tool creation, as it is one of the most essential parts of a software agent.}
{\tech} starts from a simplistic agent with only bash tool access (e.g., mini-SWE-agent~\cite{yang2024sweagent}).
During {the} regular issue-solving loop, {the agent} can synthesize, modify, and execute custom tools such as editors, code search utilities, and domain-specific analyzers.
A lightweight step‑reflection prompt repeatedly asks the agent whether creating or revising a tool would accelerate progress, thereby turning tooling into a first‑class decision alongside ordinary actions like running tests.
This mechanism leaves the underlying scaffold unchanged, requires no offline training, and is agnostic to the underlying LLM.
By elevating tool creation to an explicit, iterative decision point, we unlock this hidden \emph{on-the-fly} self-improvement ability.

\livesweagent addresses the limitations of existing research.
First, with tool creation, the agent action space adapts to the problem at hand, producing task-relevant tools that precisely capture the needs to accomplish the current task.
Furthermore, by shifting improvement from offline training to online evolution, \livesweagent alleviates tedious scaffold engineering because new abilities emerge from the encountered issues themselves.
Importantly, tool synthesis is not a one-shot preprocessing action but a continuous iterative process interleaved with problem solving. The agent can refine tools as their understanding of the failure mode evolves.
Despite its minimal and simplistic design, \livesweagent is generalizable to different agent scaffolds and LLMs, with state-of-the-art open results on software issue solving.

We have evaluated \livesweagent on the widely used \sbv benchmark~\cite{sbv} and the more challenging \swebenchpro~\cite{deng2025swe}. Without any test‑time scaling, \livesweagent achieves \textbf{77.4\%} resolve rate on Verified and \textbf{45.8\%} on Pro, surpassing state-of-the-art open‑source baselines and{ even} the best commercial agent systems.
Our ablations and tooling analyses reveal that (1) custom tool creation materially improves effectiveness (higher solve rates) with minimal overhead,
(2) benefits persist across different state-of-the-art \llm backends, with better results on stronger models, demonstrating its promising future as LLMs' capabilities rapidly evolve,
and (3) synthesized tools include both generic utilities and task-aligned specializations that benefit issue-specific problem-solving.

In summary, we make the following contributions:

\begin{itemize}[leftmargin=*]
\item \textbf{The first live software agent.}  We present \livesweagent, the first \emph{live} software agent that can autonomously self-evolve its own scaffold implementation on the fly when solving real-world issues,
without any offline training or extra pipelines.
\item \textbf{Minimal and general implementation.} Our current implementation adopts a minimal and general design, where the agent starts from a minimal bash-only scaffold (e.g., \minisweagent) and self-improves on the fly by creating general or customized tools. The design is compatible with any existing software agent loop or LLM with negligible overhead, has been publicly available at: {\color{purple}\url{https://github.com/OpenAutoCoder/live-swe-agent}}.
\item \textbf{State-of-the-art performance.}
On \sbv and \swebenchpro, \livesweagent reaches 77.4\% and 45.8\% solve rates (without any test-time scaling), respectively, outperforming all existing open-source agents and commercial systems at the time of writing. To our knowledge, our \swebenchpro result is also the best reported to date.
\item \textbf{Comprehensive analysis.} We include detailed studies of when and why on-the-fly tool creation helps, how it improves efficiency, and how it compares to other designs like fixed-tool agents, workflow-based systems, and offline self-improving methods. Notably, compared to existing self-improving agents, we achieve much better performance (65.0\% solve rate on the \sbv-60 subset vs. \dgm's 53.3\%)
while being significantly less costly (no offline cost).

\item \textbf{Unified leaderboard.} For software tasks, recent LLMs are often benchmarked using manually engineered, proprietary agent scaffolds, making fair model comparison hard. 
 \livesweagent offers an open, unified, and powerful scaffold that enables genuinely fair, apples-to-apples comparison for future model releases. We are maintaining a leaderboard of recent models evaluated on real-world software tasks using \livesweagent at: {\color{purple}\url{http://live-swe-agent.github.io}}
\end{itemize}

\section{Approach}

\begin{figure}
    \centering
    \includegraphics[width=1\columnwidth]{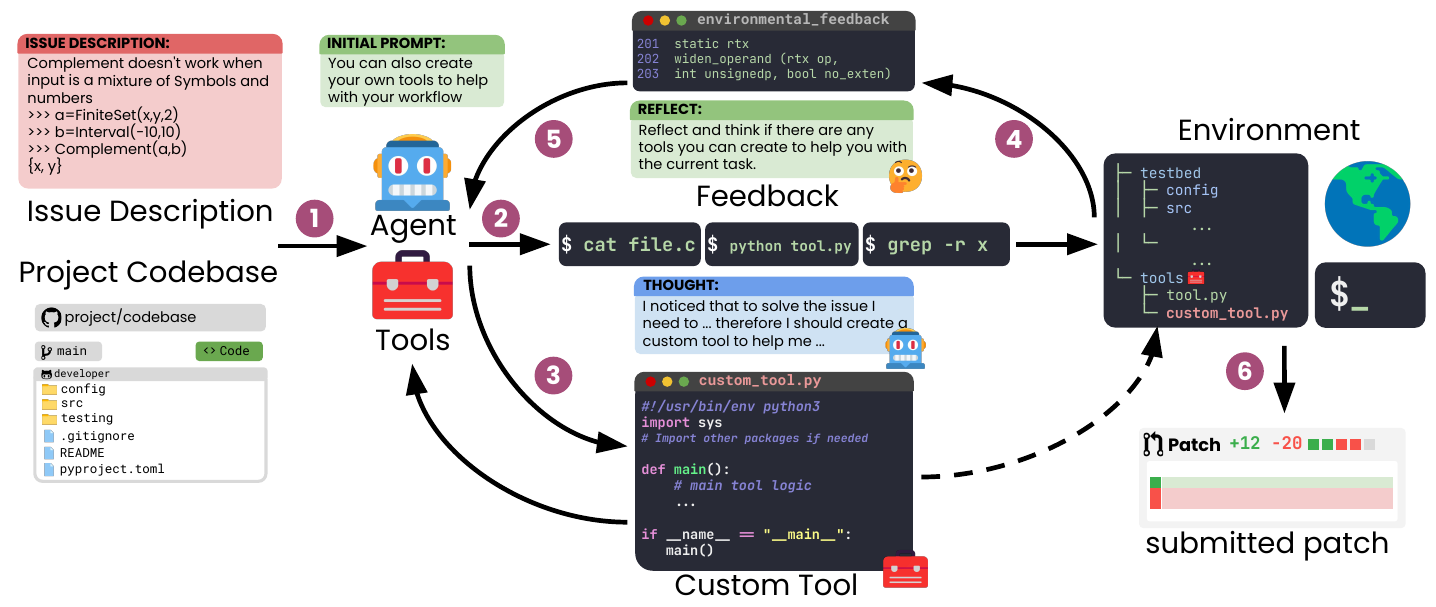}
    \caption{Overview of \tech}
    \label{fig:overview}
\end{figure}

\tech is a live, self-evolving agent that improves and expands its capabilities \textit{on the fly} while solving an issue. 
Our key insight is recognizing that agents themselves can be iteratively improved, just like the software issues they are designed to solve.
The space in which an agent can evolve includes not only the tools it uses but also the underlying agent scaffold itself.
In this work, we start from a simple agent scaffold and focus primarily on self-evolution for tool creation and usage, as tools represent one of the most critical components of an agent.
We now describe how \tech provides a framework for \llm{s} to develop and use their own tools on the fly.

Figure~\ref{fig:overview} presents an overview of \tech{}.
First, \circled{1} we take in both the project codebase and the description of the issue to be solved. 
We then provide this information to the agent and initialize it with a set of tools.
In the beginning, the agent may only have access to a limited set of tools (e.g., bash commands) and the goal of \tech is to allow the agent to generate and use its own tools \textit{on the fly} while solving the issue.
As the agent is solving the issue, at each step,{it} can choose to either \circled{2} output a command (e.g., to use a tool) or \circled{3} create a custom tool that can help it solve the issue.
In \tech, we define a custom tool as a script that can be executed in the environment. 
This provides an intuitive and straightforward tool-usage interface for the agent, allowing it to output a command to use any custom tool.
Next, different from existing approaches that \circled{4} directly provide the environmental feedback output to the agent, \circled{5} we specifically ask the agent to reflect upon the past steps and decide whether a tool should be created in the feedback message.
This loop is repeated until the agent has submitted a solution \circled{6} to the initial problem.
Unlike prior agentic setups where the set of tools and possible actions available to the agent is fixed, \tech allows the agent to perform live self-evolution by creating and using custom tools on the fly.

\subsection{On-the-fly Self Evolution}

The key idea of \tech is to enable the agent self-improve by modifying its{ own scaffold}{, such as} creating custom tools based on the problem and previous trajectories.
To support tool creation, we apply several simple modifications to the initial prompt of the agent.
Specifically, we provide instructions and examples illustrating how a tool should be created and used.
More importantly, we indicate to the agent that: (1) the goal of any tools it creates should be to help it better solve the tasks and (2) the created tools can be for any purpose and do not need to be general. 

In addition to introducing the ability for agents to create tools through the initial prompt, we also explicitly ask the agent to reflect on its past trajectory to determine if it should create any tools after each step. 
This is done by appending a simple reflection message after each environmental feedback.
In our experiments, we found that this reflection process is necessary to remind the agents to design tools that are useful and specific for the particular issue.

It is important to point out that the modifications made to a regular agentic framework by \tech are extremely simple (See Appendix~\ref{appendix:initial_prompt} for both the initial prompt and reflection message).
\tech does not make any changes to the agentic loop, impose a particular workflow, or require any costly offline training.
Instead, the focus is on allowing and extending the ability for agents to create their own custom tools at runtime to improve performance and reduce manual tool-creation effort.
This allows \tech to be extremely general and applicable across a wide-range of different tasks, \llm{s}, and agent scaffolds. 
Furthermore, we also note that software agents, in essence, are also software.
As such, they can be modified and updated on the fly by software agents (themselves) no different than any other software repository.
In \tech, we leverage this insight for on-the-fly self-improvement by enabling the agents to create their own custom tools depending on the problem at hand.
We next describe the custom tools agents create in more detail.

\subsection{Custom Tool Synthesis}

In \tech, we define a custom tool as a script that can be executed in the environment.
This allows the agents to both easily create tools (by creating a script file) and use the created tools (by running the script with arguments).
We believe this is a general and intuitive interface that is suitable for a wide range of tasks.

\begin{wrapfigure}{r}{0.4\textwidth}
  \begin{minipage}{\linewidth}
  \centering
  \captionsetup[subfigure]{justification=centering}
  \includegraphics[width=\linewidth]{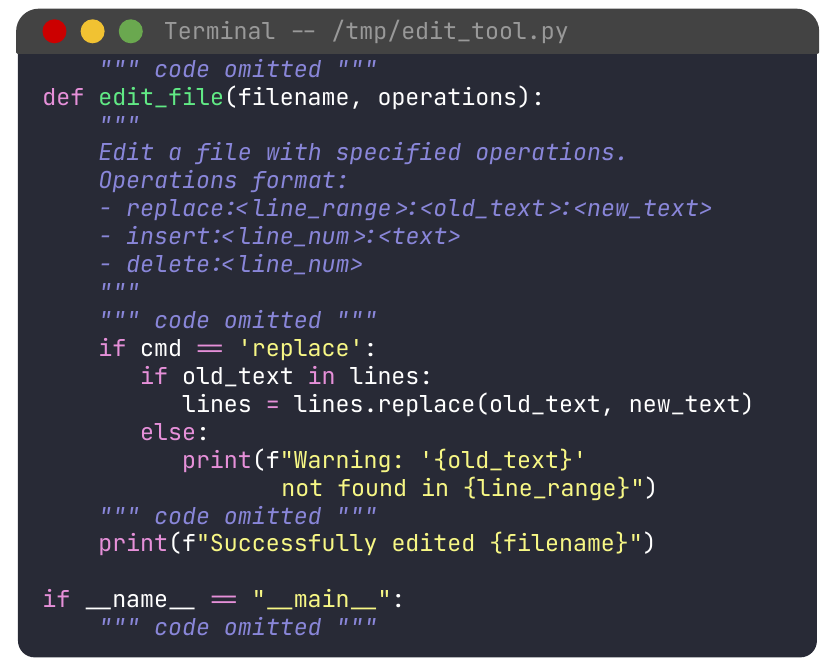}
  \subcaption{Edit tool}
  \label{fig:example_tools}
  \includegraphics[width=\linewidth]{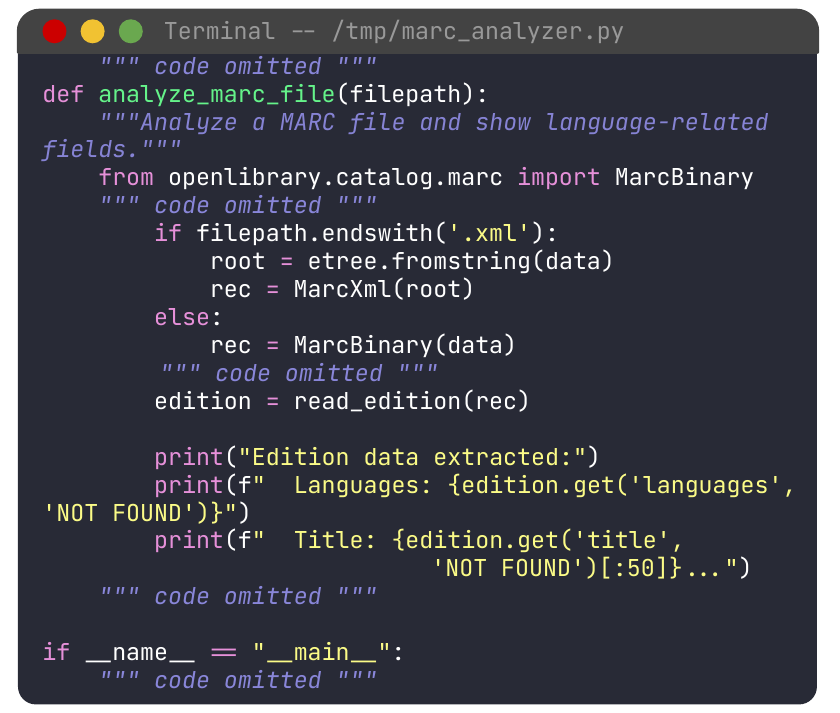}
  \subcaption{\CodeIn{MARC} file analyzer tool}
  \label{fig:example_tools2}
  \end{minipage}
  \caption{Example custom tools}
  \label{fig:example_tool_large_1}
\end{wrapfigure}

\parabf{Example editing tool.} Figure~\ref{fig:example_tools} shows an example of a custom tool created by the agent. 
This is an editing tool that allows the agent to edit a file by replacing, inserting, or deleting code.
We see that it contains the necessary tool logic as well as clear instructions on how the tool can be used.
Compared with bash commands which can potentially overwhelm the agent with many different arguments and flags, the tools that agents create themselves have a straightforward purpose and are easy to use.
Furthermore, in the example, we see that the editing tool also provides relevant feedback messages such as indicating whether the replacement edit was successful or not.
This feedback can be critically important to inform the next actions taken by the agents.
On the other hand, a bash editing command like \CodeIn{sed} does not indicate the result of the edit such that a replacement operation where the string to be replaced does not exist will produce no warning message but still return a success code.
As such, the agent may be misguided into thinking that the edit was successful when, in fact, no changes have been made to the file.

\tech also encourages agents to create tools that can improve the efficiency of their workflows. 
For example, a common step in solving an issue is to identify the locations of relevant code snippets in order to understand the root cause.
In this case, a custom search tool can be useful for searching code within certain directories and displaying their surrounding context.
While it is possible to achieve the same outcome of a custom search tool using a combination of different bash commands (e.g., \CodeIn{grep}, \CodeIn{find}, \CodeIn{cat}), the agent often needs to take multiple steps to perform the actual search, leading to increased context length and time required to solve an issue.
By creating custom tools that can handle complex and multi-step tasks, \tech can improve both the effectiveness and efficiency of agents.

\textbf{Example issue-specific tool.}
In addition to general tools, another benefit of \tech is its ability to create issue-specific tools.
Figure~\ref{fig:example_tools2} shows an example of a custom tool that is specifically tailored to a particular issue.
In this example, the tool analyzes \CodeIn{MARC} files, a file format for publication or text records, and prints them in a human-readable format. 
The agent can create this tool and use it to display the content of relevant \CodeIn{MARC} files (including binary files) that serve as test cases, helping it better understand the issue and evaluate potential patches.
Such functionality cannot be achieved easily using simple bash commands or even general-purpose tools.
By enabling the agents to create arbitrary custom tools on the fly, \tech can generate specialized tools for individual problems, allowing it to solve issues more effectively.

A fair question to ask is \textit{why not generate these tools directly in the beginning?} 
The reason is that tool creation, similar to manual problem solving, is also an iterative process.
We need to understand the problem at hand and during the solving process identify the issues in order to come up with helpful tools in different scenarios.
For example, in the \CodeIn{MARC} file example, it is not immediately obvious that we need to create a custom tool specifically to analyze \CodeIn{MARC} files.
By creating all the tools in the beginning and applying this fixed set of tools during the entire process, we will lose the opportunity to design custom tools that are helpful in unique situations.
Additionally, having access to all possible tools are not necessarily better as they can overwhelm and mislead the agent.
Furthermore, the agent may often iterate on or modify the tools it has created which requires runtime modification abilities not available under a fixed tool setup. 
\tech provides the ability for agents to automatically synthesize customs tools on the fly without any additional overhead from heavy scaffold modifications or costly offline training updates.

\section{Experimental Setup}
\label{sec:setup}

\parabf{Implementation.} 
While \tech is general across different software agent scaffolds, we implement \tech on top of the popular \minisweagent~\citep{yang2024sweagent} framework. 
As such, we retain the hyperparameters used in \minisweagent by default (i.e., maximum step limit of 250 and maximum cost of \$3 per issue).
We choose \minisweagent to build on since it is not only simplistic (with $\sim$100 lines of code and only accessing bash commands) but also widely used.
Unless otherwise stated, we use \claudesonnetfourfive (\CodeIn{claude-sonnet-4-5-20250929})~\citep{sonnet4.5} in our experiments and sample one patch per issue.

\parabf{Datasets.}
We evaluate \tech on the popular \swebench Verified benchmark~\citep{sbv} containing 500 software development problems where the goal is to successfully modify the repository given a problem description. 
\swebench Verified is validated by human developers to ensure each problem description has sufficient amount of information to solve the issue.
Additionally, we also evaluate on the recent \swebenchpro~\citep{deng2025swe} benchmark, containing 731 publicly available problems, aimed to capture realistic, complex, and enterprise-level problems. 
Compared with \swebench Verified, \swebenchpro contains more difficult problems across multiple repositories and programming languages.

\parabf{Baselines.}
We compare \tech against representative state-of-the-art agentic approaches. 
For \swebench Verified, we compare against \minisweagent as it is one of the most widely-used open-source agentic solutions on \swebench tasks with top leaderboard performance.
Additionally, it is also a straightforward comparison as we directly build on top of the \minisweagent framework.
Furthermore, we compare with prior self-evolving agents: Self-Improving Coding Agent (SICA)~\citep{robeyns2025sica}, Darwin-Gödel Machine (DGM)~\citep{zhang2025darwin}, and Huxley-Gödel Machine (HGM)~\citep{wang2025huxley} on a subset of \swebench Verified problems. 
This subset of 60 \swebench Verified problems has been used by prior work~\citep{wang2025huxley} to specifically evaluate all three self-improving agent baselines.
For \swebenchpro, we compare against \sweagent~\citep{yang2024sweagent} which is the top-performing approach on the \swebenchpro leaderboard.
For each baseline, we directly reuse their experimental results and report their performance, cost, as well as the backend \llm used whenever possible.

\section{Evaluation}

\begin{table}[t!]
\centering
\caption{Result on \swebench{} Verified}
\label{tab:verified}
\scalebox{1}{
\begin{tabular}{ll|rr}
\toprule
 Tool & \llm & {\%{} Resolved} & {Avg. \$ Cost}\\
\midrule
 & \openai{} \gptfivemini{} & 59.8\% & \$0.04 \\
 & \openai{} \gptfive{} & 65.0\% & \$0.28 \\
 & \anthropic{} \claudesonnetfourfive{} & 70.6\% & \$0.56 \\
 \multirow{-4}*{\minisweagent{}~\citep{yang2024sweagent}} & \gemini{} \geminithreepro{} & 74.2\% & \$0.46 \\
\midrule
 & \openai{} \gptfivemini{} & 63.0\% & \$0.05 \\
 & \openai{} \gptfive{} & 68.4\% & \$0.27 \\
 & \anthropic{} \claudesonnetfourfive{} & 75.4\% & \$0.68 \\
\multirow{-4}*{\textbf{\tech{}}} & \gemini{} \geminithreepro{} & 77.4\% & \$0.48 \\
\bottomrule
\end{tabular}
}
\end{table}

\begin{table}[t!]
\centering
\caption{Result on \swebench{} Verified-60{}}
\label{tab:verified_subset}
\scalebox{1}{
\begin{tabular}{lll|rr}
\toprule
& Tool & \llm & {\%{} Resolved} & {Offline cost (hours)}\\
\midrule
\multirow{3}{*}{\rotatebox[origin=c]{90}{\makecell{\footnotesize{}Offline self-\\improving\\agents}}} & \sica{}~\citep{robeyns2025sica} & \openai{} \gptfivemini{} & 50.0\% & infinite loop \\
\cmidrule{2-5}
& \dgm{}~\citep{zhang2025darwin} & \openai{} \gptfivemini{} & 53.3\% & 1231 \\
\cmidrule{2-5}
& \hgm{}~\citep{wang2025huxley} & \openai{} \gptfivemini{} & 56.7\% & 512 \\
\midrule
\midrule
& {\textbf{\tech{}}} & \openai{} \gptfivemini{} & 65.0\% & 0 \\
\bottomrule
\end{tabular}
}
\end{table}

\subsection{Main Results}

\parabf{\swebench{} Verified.} 
Table~\ref{tab:verified} shows the results of \tech and prior agent approaches on \swebench Verified. 
We first observe that compared with \minisweagent{}, across four different \llm backends, \tech consistently achieves a higher resolve rate.
This demonstrates the improvement in performance enabled by allowing agents to create and use their own custom tools on the fly.
Furthermore, \tech is able to achieve this with only a minimal increase in cost compared with the base \minisweagent.
In certain cases (e.g., in \gptfive{}), we even observed slight cost savings where the agent can improve the solving efficiency by replacing complex, multi-turn commands with customs tools that achieve the same functionality.

We also demonstrate the comparison between \tech and the best-performing agentic tools, including both state-of-the-art open-source solutions and proprietary commercial scaffolds, on \swebench Verified (Figure~\ref{fig:result}).
We only report single-attempt results in Figure~\ref{fig:result} without any test-time scaling to ensure a fair comparison.
We observe that \tech using \geminithreepro{} achieves \textbf{a solve rate of 77.4\% without test-time scaling, outperforming all existing agents on the \swebench Verified leaderboard}\footnote{\url{https://www.swebench.com}} including state-of-the-art commercial solutions at the time of writing.

Furthermore, we perform a detailed comparison with three prior self-evolving agents on a subset of 60 \swebench Verified problems chosen by prior evaluation~\citep{wang2025huxley}.
Table~\ref{tab:verified_subset} shows the resolve rate and the offline cost for \sica, \dgm, \hgm, and \tech on the subset of problems.
We observe that \tech achieves the best performance, improving the resolve rate by 8.3 percentage points compared to previous best approach.
We also see that previous self-evolving agents all require a heavy amount of offline training to evolve the base agent, costing more than 500 hours.
Furthermore, prior self-evolving techniques produce a static agent used for all problems.
On the other hand, \tech creates customs tools for each individual task, allowing it to adapt on the fly based on the problem and specific LLM used.
Unlike the costly offline updates required by prior self-evolving agents, \tech instead adopts an online evolving approach to prompt the agent to generate custom tools on the fly to improve performance with minimal overhead.

\begin{table}[t!]
\centering
\caption{Result on \swebenchpro{}}
\label{tab:pro}
\scalebox{1}{
\begin{tabular}{ll|rr}
\toprule
 Tool & \llm & {\%{} Resolved} & {Avg. \$ Cost}\\
\midrule
\sweagent{}~\citep{yang2024sweagent} & \anthropic{} \claudesonnetfourfive{} & 43.6\% & - \\
\midrule
\textbf{\tech{}} & \anthropic{} \claudesonnetfourfive{} & 45.8\% & \$0.73 \\
\bottomrule
\end{tabular}
}
\end{table}

\parabf{\swebenchpro{}.} 
We also evaluate \tech on the public set of \swebenchpro, containing 731 problems across 11 repositories and four programming languages (Python, Go, TypeScript, and JavaScript).
Table~\ref{tab:pro} shows the performance of \tech compared to the state-of-the-art baseline of \sweagent on \swebenchpro~\cite{deng2025swe}.
Different from \minisweagent, which only has access to basic bash commands, \sweagent provides handcrafted file viewing and editing tools and is the best performing approach on the \swebenchpro.
We choose \claudesonnetfourfive{} for \sweagent because it ranks at the top of \swebenchpro leaderboard\footnote{\url{https://scale.com/leaderboard/swe_bench_pro_public}}, as such, we also use it as the base \llm for \tech to enable a fair comparison.
We observe that \tech is able to achieve better performance compared with \sweagent, a specially designed agent implemented with close to 7,000 lines of code.
We also compare with all other top-performing agents and \llm{s} in Figure~\ref{fig:result} and observe that \tech is able to achieve the \textbf{new state-of-the-art performance of 45.8\% resolve rate on \swebenchpro}.
This further demonstrates the superiority of the live scaffold design of \tech compared to existing agents that interact with a fixed set of manually-crafted tools.

\subsection{Tools Analysis}
\label{sec:tool_analysis}

\begin{figure*}[h]
    \centering
    \begin{subfigure}[b]{0.32\textwidth}
    \centering
    \includegraphics[width=\textwidth]{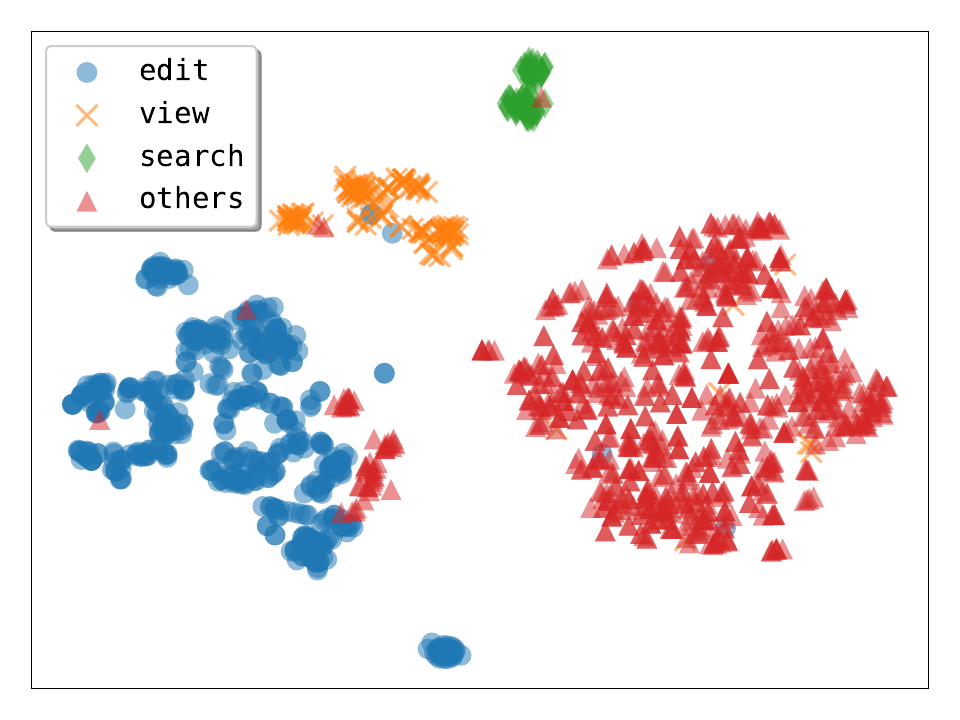}
    \caption{Tools in \swebench Verified}
    \label{fig:tool_verified}
\end{subfigure}
\begin{subfigure}[b]{0.32\textwidth}
    \centering
    \includegraphics[width=\textwidth]{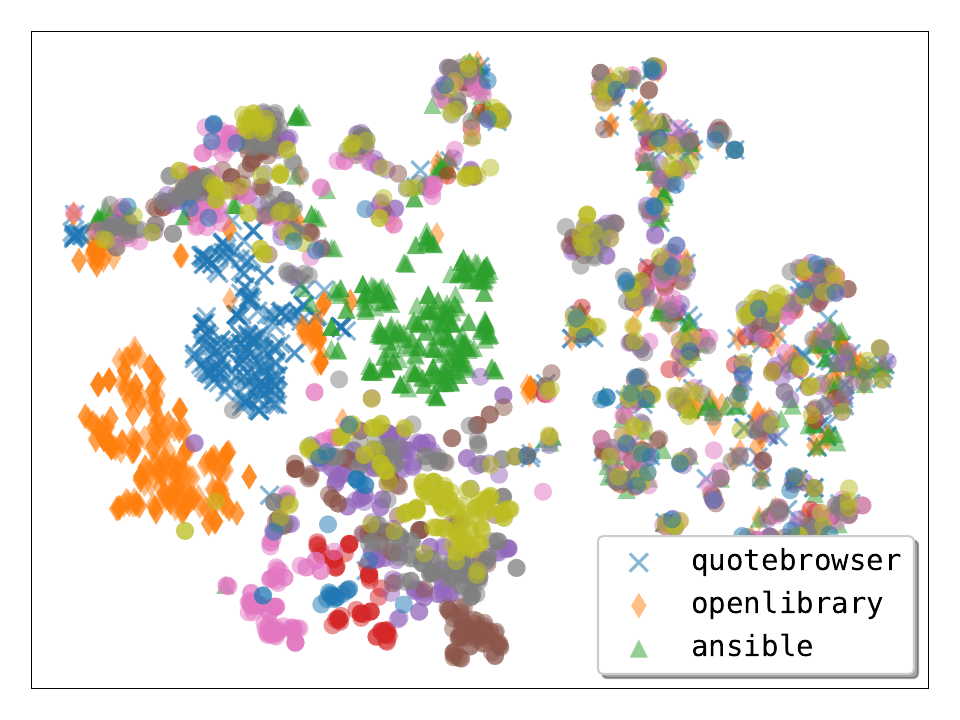}
    \caption{Tools in \swebenchpro{}}
    \label{fig:tool_pro_repo}
\end{subfigure}
\begin{subfigure}[b]{0.32\textwidth}
    \centering
    \includegraphics[width=\textwidth]{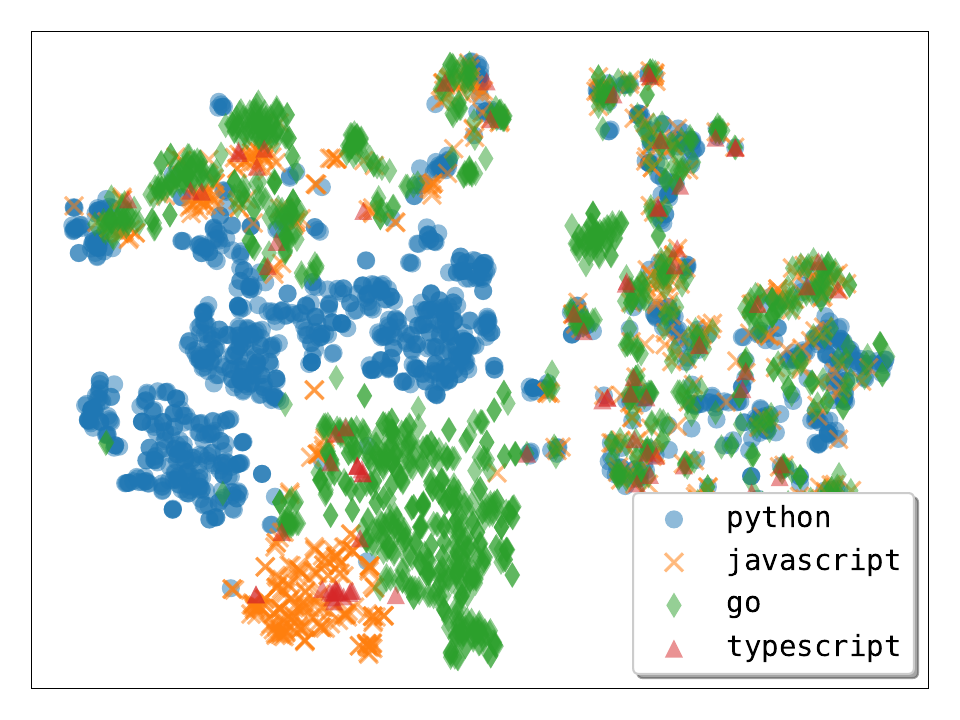}
    \caption{Tools in \swebenchpro{}}
    \label{fig:tool_pro_language}
\end{subfigure}
\caption{
2 dimensional t-SNE visualization of tools generated by \claudesonnetfourfive{} on \swebench Verified and \swebenchpro.
We label and display the embedding based on tool type (Figure~\ref{fig:tool_verified}), repository name (Figure~\ref{fig:tool_pro_repo}), and programming language used in the repository (Figure~\ref{fig:tool_pro_language}).
Note that for Figure~\ref{fig:tool_pro_repo}, we only label three repositories in the legend due to space considerations. 
The three repositories were chosen as they have representative distinct clusters.
}

\label{fig:tsne}
\end{figure*}

\parabf{Categories and variations of custom tools.}
We examine the custom tools created by \tech{}. 
Figure~\ref{fig:tsne} shows the embedding visualizations using t-SNE~\citep{hinton2002stochastic} of the tools created by \tech{}. 
We compute the embedding for each tool based on the tool body (i.e., the content of the tool script) using a text embedding model (OpenAI \CodeIn{text-embedding-3-small}).
To start off, we look at the types of tools created by \tech by categorizing them into common tool functionalities (e.g., edit, view, search, etc). 
We perform this categorization by using simple string matching based on the script filename of the tool.
Figure~\ref{fig:tool_verified} shows the visualization of the tools generated for \swebench Verified. 
We see that while there are definitely distinct clusters of common tools like edit, view, and search, there are still variations among them.
For example, the edit tool cluster (highlighted in blue circles) is not a tightly concentrated dot, but instead is spreadout, showing that \tech can generate different editing tools depending on the problem.
Furthermore, \tech also allows the agent to generate additional tools (the others category highlighted in red triangles). 
These are the more unique issue-specific tools such as a script that applies a very detailed patch to multiple files or a diff checking tool that shows the difference between two files.
We also apply similar analysis to all studied datasets in Appendix~\ref{appendix:tool_analysis}.

In addition to the types of tools created, we also examine how the tools created can change across different repositories and languages.
Figure~\ref{fig:tool_pro_repo} shows the visualization of the tools created in \swebenchpro categorized based on repositories.
In \swebenchpro, there are 11 different repositories, and not surprisingly, there are common tools created across different repositories (i.e., the overlapping clusters in the figure). 
On the other hand, there are also specific tools for certain repositories, for example we see a distinct cluster (highlighted in orange diamonds) for \CodeIn{openlibrary}, a codebase for cataloging literature.
Since \CodeIn{openlibrary} deals with lots of raw data of a specialized format, the custom tools generated by \tech are specifically tailored for that.
We also observe similar results when we break down the tools based on programming language of the repositories in Figure~\ref{fig:tool_pro_language}.
Different problems, environments, and languages may require completely different specialized tools, highlighting again how using the same set of tools for all problems is suboptimal.

\begin{wrapfigure}{r}{0.4\textwidth}
  \begin{minipage}{\linewidth}
  \centering
  \captionsetup[subfigure]{justification=centering}
  \includegraphics[width=\linewidth]{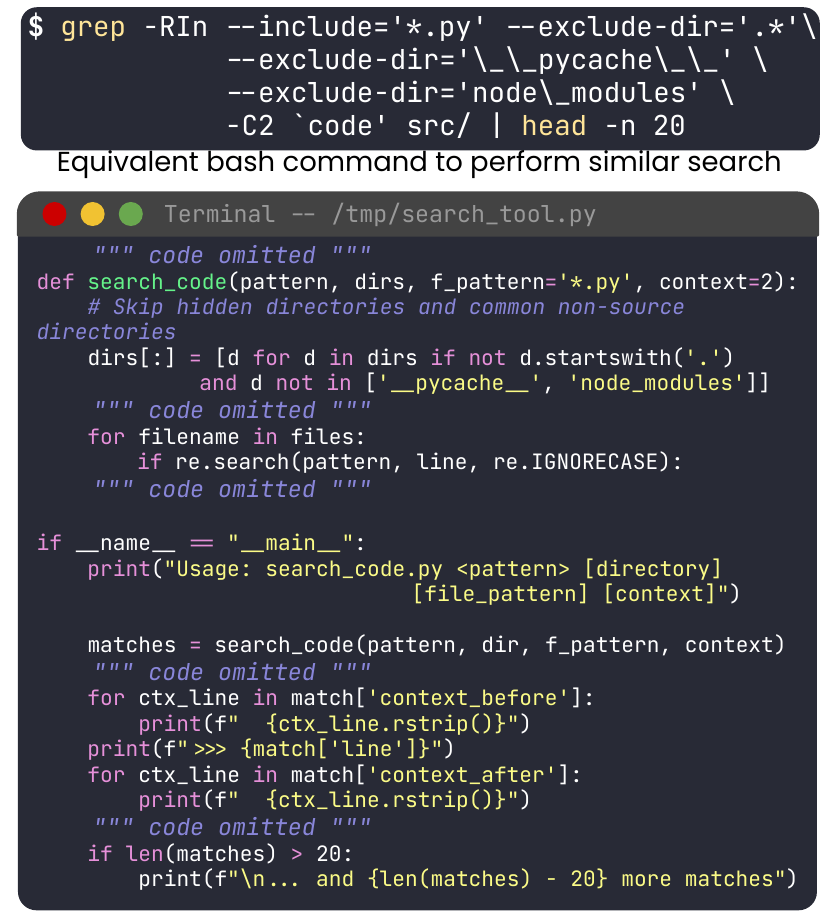}
  \subcaption{Search tool}
  \label{fig:example_tools3}
  \includegraphics[width=\linewidth]{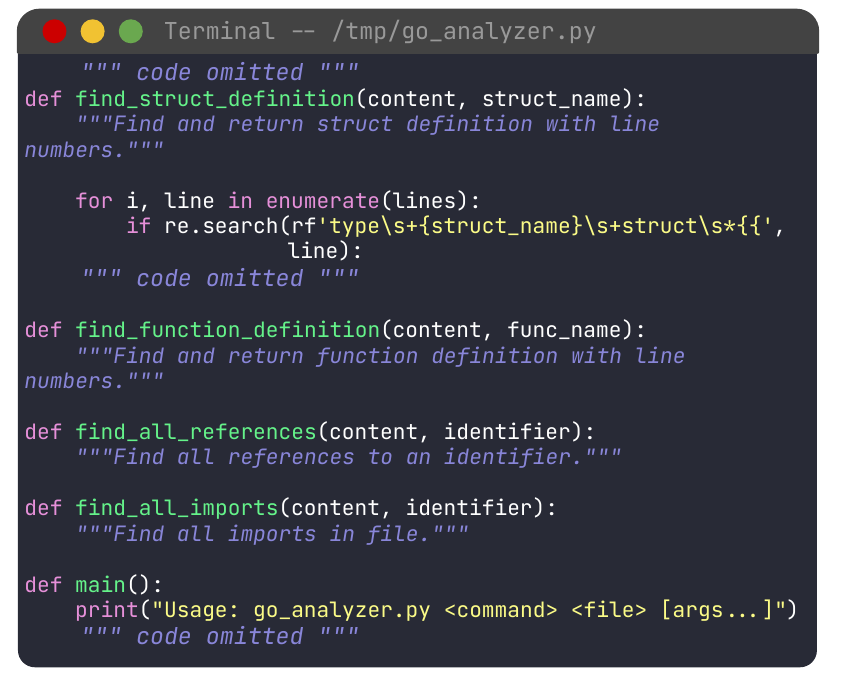}
  \subcaption{Go file analyzer tool}
  \label{fig:example_tools4}
  \end{minipage}
  \caption{Example custom tools}
  \label{fig:example_tools_big_2}
\end{wrapfigure}

\parabf{Effective and interesting custom tools.}
We now take a closer look at a few exemplary effective and interesting tools created by \tech.
Figure~\ref{fig:example_tools3} shows a custom search tool created by \tech.
This example might appear simple and straightforward at first glance.
However, it implements several important features: (1) supports focused searching within directories and pattern-based matching, (2) ignores irrelevant folders (\CodeIn{\_\_pycache\_\_} and \CodeIn{node\_modules}) and hidden folders (\CodeIn{.git}) during search, (3) displays relevant code context before and after the found code, and (4) limits results to top 20 matches to reduce context size.
By baking in the exclusion of irrelevant folders in the search tool, agents do not have to add these folders as a separate flag during the searching process.
Showing the relevant surrounding context is also helpful since oftentimes we want to find out how the code snippet of interest is being used.
Furthermore, only showing the first 20 matches can be critically important to not drastically inflate the context window, allowing the agent to easily refine their search to limit results.
An example bash command to emulate the functionality of the custom tool is shown at the top of Figure~\ref{fig:example_tools3}.
We can see the high degree of complexity in terms of options, arguments, and flags used in this long command.
On the other hand, we can simply call \CodeIn{"python search\_code.py `code' src/"} to perform the search via the custom tool.
By only using basic bash commands, the agent will have to chain together multiple commands using complex flags with potential to incur errors (e.g., older \CodeIn{grep} versions do not support features like \CodeIn{--exclude-dir}) or substantial slowdowns.
In fact, in our experimental results, we found that without access to custom tools, the agent may often take multiple steps to accomplish a basic task such as searching for relevant code.
This issue is further amplified when steps like search may need to be done multiple times for one task thus drastically limiting the problem solving efficiency and effectiveness.
By creating and introducing efficient and easy to use custom tools designed by agents themselves, \tech can improve agent performance on complex software engineering tasks.

Figure~\ref{fig:example_tools4} shows another custom tool \CodeIn{go\_analyzer.py} created by \tech. 
The tool can be used to analyze a Go file to find any struct and function definition, identifier references, and to obtain the imports used in the file.
This example showcases the powerful generative capability of agents where the custom tool can be used as a simple static analyzer for the Go programming language to search and provide key information about a file.
Different from the previous search tool example where one can technically mimic the behavior using bash commmands, abeit using very complex chain commands, the tool logic here is even more complex.
The custom tool performs strategic pattern matching based on Go grammar and provides an interface to support multiple different usages.
Furthermore, by saving this custom tool in an executable script, the agent can reuse the functionality multiple times.
By creating and using this custom tool to better understand file structure and content, \tech is able to solve this issue\footnote{\CodeIn{navidrome\_\_navidrome-10108c63c9b5bdf2966ffb3239bbfd89683e37b7}, \swebenchpro{}} that prior best performing baseline could not.

\subsection{Ablation}
\label{sec:ablation}

To evaluate the effect of different setups of \tech, we use a random subset of 50 problems in \swebench Verified (see Appendix~\ref{appendix:ablation_dataset} for the detailed list of problems).
Unless otherwise stated, we use \claudesonnetfourfive{} in our ablation experiments.

\begin{table}[t!]
\centering
\caption{Ablation result across \tech setups on subset of \swebench Verified problems}
\label{tab:ablation_setup}
\scalebox{1}{
\begin{tabular}{l|ll}
\toprule
 Approach & \% Resolved & \# of tools created\\
 \midrule
\tech{} w/o tool creation & 62.0\% & 0.00 \\
\tech{} w/o reflection & 64.0\% & 2.92 \\
\tech{} & 76.0\% & 3.28 \\
\bottomrule
\end{tabular}
}
\end{table}

\parabf{Effectiveness of tool creation steps.} 
Table~\ref{tab:ablation_setup} shows the impact of different components of \tech have on performance.
We see that by removing the on-the-fly tool creation ability of \tech (i.e., using the base \minisweagent), we achieve the lowest solve rate.
The performance is improved when we indicate to the agent to create custom tools in the initial prompt (row \tech w/o reflection).
However, the highest resolve rate is achieved once we explicitly ask the agent to reflect on the past trajectories to determine if they should create any tools after each step.
In our experiments, we found that this reflection process provides a good reminder for the agent to create tools that are specifically designed for the particular issue.
Furthermore, while the number of tools do not necessary correlate with solving performance, we also see that this reflection process creates more tools on average compared to only indicating the tool creation in the initial prompt. 
\tech is able to improve performance by reflecting on both the current problem and past trajectories to create custom tools on the fly.

\begin{table}[t!]
\centering
\caption{Ablation result across different \llm backends on subset of \swebench Verified problems}
\label{tab:ablation_llm}
\scalebox{1}{
\begin{tabular}{l|ll}
\toprule
 \llm & \minisweagent & \tech{}\\
 \midrule
\openai{} \gptfivenano{} & 44.0\% & 14.0\% \textcolor{red}{($\downarrow$-68.2\%)} \\
\openai{} \gptfivemini{} & 60.0\% & 58.0\% \textcolor{red}{($\downarrow$-3.3\%)} \\
\openai{} \gptfive{} & 60.0\% & 68.0\% \color[HTML]{009933}{($\uparrow$13.3\%)} \\
\midrule
\anthropic{} \claudesonnetthreeseven{} & 46.0\% & 50.0\% \color[HTML]{009933}{($\uparrow$8.7\%)} \\
\anthropic{} \claudesonnetfour{} & 58.0\% & 64.0\% \color[HTML]{009933}{($\uparrow$10.3\%)} \\
\anthropic{} \claudesonnetfourfive{} & 62.0\% & 76.0\% \color[HTML]{009933}{($\uparrow$22.6\%)} \\
\bottomrule
\end{tabular}
}
\end{table}

\parabf{Different \llm backends.} 
We now examine the effect of using different \llm backends for \tech.
Table~\ref{tab:ablation_llm} compares the performance of the base \minisweagent and \tech as we vary the \llm used on the same 50 problems used in the previous ablation experiment.
We first observe that there are weaker \llm{s} for which using \tech to create custom tools on the fly even decreases performance.
In particular, we see that for \gptfivenano{}, the results are significantly worse when using \tech compared to base \minisweagent.
After examining the trajectories, we found that \gptfivenano{} fails to understand the goal of creating custom tools and is often stuck in a loop thus leading to reduction in performance.
This indicates that weaker \llm{s} may lack the high-level reasoning capabilities to synthesize useful tools on the fly.
However, we see that \tech is able to improve more over the base \minisweagent as we start to use more powerful models.
Specifically, state-of-the-art \llm{s} such as \claudesonnetfour{} and \gptfive{} achieves the highest relative resolve rate improvement when using \tech compared with \minisweagent.
This demonstrates the generalizability of \tech across high-performing \llm{s} and the potential for \tech to further improve performance especially as we continue to build more and more powerful \llm{s}.

\begin{table}[t!]
\centering
\caption{Result on a subset of \swebenchmultilingual{} problems}
\label{tab:multilingual}
\scalebox{1}{
\begin{tabular}{ll|rr}
\toprule
 Tool & \llm & {\%{} Resolved} & {Avg. \$ Cost}\\
\midrule
\minisweagent{}~\citep{yang2024sweagent} & \anthropic{} \claudesonnetfourfive{} & 40.0\% & \$0.59 \\
\midrule
\textbf{\tech{}} & \anthropic{} \claudesonnetfourfive{} & 46.0\% & \$0.66\\
\bottomrule
\end{tabular}
}
\end{table}

\parabf{\swebenchmultilingual{}.} 
In addition to evaluating on \swebench Verified and \swebenchpro, we also test the generalizability of \tech on other benchmarks. 
We conducted an initial experiment on a subset of 50 problems (see Appendix~\ref{appendix:multilingal_subset} for the detailed list of problems) in \swebenchmultilingual~\citep{sbm}, a benchmark of software engineering tasks across 9 programming languages (JavaScript, TypeScript, Rust, Ruby, Go, C/C++, PHP, and Java).
\Cref{tab:multilingual} shows the performance of \tech compared to \minisweagent on \swebenchmultilingual. 
We observed that \tech achieves a better performance by obtaining a resolve rate of 46.0\%, while \minisweagent only has a resolve rate of 40.0\%.
This demonstrates the generalizability of \tech on additional challenging problems.

\subsection{Discussion and Future Work}

\parabf{Towards general, on the fly self-evolution.}
In this work, \tech primarily focuses on enabling agents to self-evolve through custom tool creation and usage.
As discussed previously, the key idea of \tech, to improve and modify the agent itself on the fly, extends beyond creating new tools to also modifying the entire agent implementation.
This includes the overall system prompt of the agent, how the agent interacts with the environment, and even the concrete workflow of the agent when it attempts to solve the issue.
We believe that the entire agentic loop, much like any other software, can be modified and improved especially on the fly based on feedback and insights gained from solving a problem.
Furthermore, we hope to extend the self-evolution loop across different tasks as well.
Instead of discarding each evolved agent after a task is completed, we can save and
serialize the useful tools and insights (via concepts like Skills~\cite{skills}) for future tasks.
The agent can then load these useful tools and insights obtained while solving previous tasks on the fly  
to further improve performance and support continuous self-evolution across tasks.

\parabf{Impact on \llm evaluation.}
Our experimental results on multiple benchmarks using multiple \llm{s} demonstrate that \tech can achieve state-of-the-art performance.
This shows that \tech can provide an effective scaffold for unified evaluation of \llm performance on solving software development issues.
Instead of using complex or proprietary agent scaffolds, \tech provides a simplistic and lightweight approach that can be easily added on top of any agent design to evaluate \llm{s}. 
Moreover, \tech not only evaluates the issue solving ability, but it can also test the tool-creation capability of \llm{s} and agents.
Tools are one of the most important aspect of an agent and directly impacts the issue solving performance.
The ability for \llm{s} to create custom tools, especially for solving complex software development issues, have not been extensively evaluated.
To this end, we have already seen some interesting results in our evaluation where weaker \llm{s} lack high-level reasoning to create useful tools on the fly (see Section~\ref{sec:ablation}).
\tech provides a unique framework to jointly evaluate both the tool creation and issue resolution ability of \llm{s}.

\parabf{Applications beyond software issue resolution tasks.}
In addition to software issue resolution tasks, \tech can be easily applied to other challenging software engineering tasks like generating tests~\cite{deng2023large}, detecting and patching vulnerabilities~\cite{lee2025sec}, and synthesizing production-ready software from scratch~\cite{zhao2024commit0}.
Compared with issue resolution tasks, other software problem domains will require even more task-specific and diverse tools and agent scaffolds.
For example, to detect malicious vulnerabilities in commercial off-the-shelf binaries, we need binary analysis tools and decompilers. 
To optimize a large complex system, we need to apply profilers and tracing tools.
Instead of using a fixed agent scaffold with basic tools that cannot adopt to different problem domains or painstakingly designing specialized agents for each individual task, \tech can easily generalize to solving tasks in different domains by automatically modifying itself on the fly based on the task at hand.

\parabf{Self-evolution during \llm training.}
In this work, \tech is implemented as a lightweight modification to an existing agent, requiring no offline training or updates.
However, the idea of on-the-fly self-evolution can also be easily extended to \llm training, where instead of learning from a fixed workflow and a static set of tools, the \llm learns to also create new tools and modify the scaffold itself during training. 
In this approach, the resulting \llm will have improved reasoning capabilities as the self-evolving training approach provides additional learning signals, allowing the \llm to solve more complex tasks. 
Through self-evolving training, \llm{s} can learn to better create useful, task-specific tools and modify advanced scaffolds dynamically based on the problem at hand. 
Moreover, the final trained \llm will be compatible with more runtime agent frameworks since it is able to learn from its own created tools and modified scaffolds rather than relying on predefined scaffolds. 
This adaptability allows it to be more robust and generalize effectively to new scaffold setups and
even different tasks.

\section{Related Work}
\subsection{Software Engineering Agents}
Inspired by the human debugging process, where developers interact with the environmental feedback (like test failures) and learn from their earlier attempts, \chatrepair~\citep{xia2023conversational,xia2024automated} proposed the first interactive bug-fixing solution based on LLMs. Since \chatrepair, a large body of research work on bug fixing and general coding tasks aims to automatically provide LLMs more context information through multi-turn conversations~\citep{chen2023teaching, kong2025contrastrepair, yuan2024evaluating}. More recently, foundation LLMs have seen substantial advances in their tool use and reasoning capabilities, and it becomes very natural to further equip feedback-driven solutions with such emergent LLM capabilities. In March 2024, Devin AI released the first AI software engineer, which can fully autonomously complete end-to-end software tasks, such as GitHub issue resolution~\citep{devin}. The initial Devin release showed an impressive resolve rate of 13.86\% on \swebench~\citep{jimenez2023swe}, a dataset with thousands of real-world GitHub issues. Since then, a large number of dedicated software agent scaffolds have been proposed, including \sweagent~\citep{yang2024swe}, \openhands~\citep{wang2024openhands}, \acr~\citep{zhang2024autocoderover}, and Trae Agent~\citep{liu2024marscode}. Such software agents typically equip the LLMs with a suite of coding tools and encourage the LLMs to autonomously decide the next actions for completing real-world software tasks. Different from the mainstream software agents, researchers have also proposed various AI software engineer solutions based on pre-defined workflows to challenge the necessity of complicated agent design, such as \agentless~\citep{xia2024agentless} and Moatless~\citep{moatless}. Moreover, recent LLMs have increasingly been post-trained on massive real-world software data to better solve software engineering issues, including SWE-RL~\cite{wei2025swerl}, DeepSWE~\cite{luodeepswe}, DeepSeek V3.1~\cite{deepseekv31}, MiniMax M1/M2~\cite{chen2025minimax}, Kimi K2~\cite{team2025kimi}, SWE-1.5~\cite{swe15}, and Code World Model (CWM)~\cite{carbonneaux2025cwm}.

Due to the huge design space for software agents, it can be extremely challenging and costly to build an optimal agent scaffold. As a result, a number of self-improving software agents have been proposed very recently, including the Self-Improving Coding Agent (SICA)~\citep{robeyns2025sica}, Darwin-Gödel Machine (DGM)~\citep{zhang2025darwin}, and Huxley-Gödel Machine (HGM)~\citep{wang2025huxley}. However, such self-improving agents require costly offline training on known benchmarks, and may not be generalize well across different LLMs, benchmarks, and issue types.
Beyond the software engineering domain, prior work~\cite{llmastool, wang2024voyager, qiu2025alitageneralistagentenabling, qian2024creatortoolcreationdisentangling, wang2024troveinducingverifiableefficient} explored using LLMs to create tools for general reasoning or embodied tasks, but they do not target real-world software engineering problems.
In contrast, in this paper, we propose \livesweagent, the first live software agent capable of performing practical self-evolution \emph{on-the-fly} during runtime when solving real-world issues. In this way, \livesweagent requires no offline training at all and can be easily generalized to different LLMs and issue domains. Moreover, it also demonstrated superior performance compared to all existing self-improving software agents.

\subsection{Benchmarks for Software Engineering Agents}

To evaluate and demonstrate the performance of software engineering agents, a large number of datasets have been proposed. \swebench~\citep{jimenez2023swe} is one of the earliest and most widely used benchmark datasets for software agents. Besides the initial \swebench dataset, which includes thousands of real-world GitHub issues, researchers have also built curated subsets with high-quality and representative issues to support faster and more reliable evaluation, including \swebench Lite~\citep{jimenez2023swe} and \swebench Verified~\citep{sbv}. Since \swebench mostly focused on Python projects, researchers have proposed a number of \swebench-style benchmarks for projects in multiple languages for more comprehensive agent evaluation, including SWE-PolyBench~\citep{rashid2025swe}, \swebench Multilingual~\citep{sbm}, and Multi-SWE-bench~\citep{zan2025multi}. Also, \swebench relies heavily on manual effort for
collecting benchmark instances and setting up executable environments, and is hardly scalable; as a result, researchers have also leveraged software agents to streamline issue collection and environment setup to build live and scalable benchmarks, including SWE-bench-Live~\citep{zhang2025swe} and SWE-rebench~\cite{badertdinov2025swe}. More recently, Scale AI has also constructed \swebenchpro~\citep{deng2025swe}, which aims to capture more realistic, complex, enterprise-level issues than SWE-bench. For rigorous evaluation of \livesweagent, our study involved multiple widely used benchmarks, including \swebench Verified, \swebench Multilingual, \swebench Pro.

\section{Conclusion}

In this paper, we have proposed \livesweagent, the first \emph{live software agent} that can autonomously and continuously evolve itself \emph{on-the-fly} during runtime when solving real-world software problems. More specifically, \livesweagent starts with the most basic agent scaffold with only access to \texttt{bash} tools, and autonomously evolves its own scaffold implementation while solving real-world software problems. 
Our evaluation on the widely studied benchmarks (such as \swebench Verified and \swebench Pro) demonstrated that \livesweagent outperforms state-of-the-art manually design software agents, demonstrating a promising future for live and self-evolving software agents.

\bibliographystyle{plain}
\bibliography{reference}

@article{carbonneaux2025cwm,
  title={Cwm: An open-weights llm for research on code generation with world models},
  author={Carbonneaux, Quentin and Cohen, Gal and Gehring, Jonas and Kahn, Jacob and Kossen, Jannik and Kreuk, Felix and McMilin, Emily and Meyer, Michel and Wei, Yuxiang and Zhang, David and others},
  journal={arXiv preprint arXiv:2510.02387},
  year={2025}
}

@article{xia2024agentless,
  title={Agentless: Demystifying llm-based software engineering agents},
  author={Xia, Chunqiu Steven and Deng, Yinlin and Dunn, Soren and Zhang, Lingming},
  journal={arXiv preprint arXiv:2407.01489},
  year={2024}
}

@article{wei2025swerl,
  title={SWE-RL: Advancing LLM Reasoning via Reinforcement Learning on Open Software Evolution}, 
  author={Yuxiang Wei and Olivier Duchenne and Jade Copet and Quentin Carbonneaux and Lingming Zhang and Daniel Fried and Gabriel Synnaeve and Rishabh Singh and Sida I. Wang},
  year={2025},
  journal={arXiv preprint arXiv:2502.18449}
}

@article{yuan2024evaluating,
  title={Evaluating and improving chatgpt for unit test generation},
  author={Yuan, Zhiqiang and Liu, Mingwei and Ding, Shiji and Wang, Kaixin and Chen, Yixuan and Peng, Xin and Lou, Yiling},
  journal={Proceedings of the ACM on Software Engineering},
  volume={1},
  number={FSE},
  pages={1703--1726},
  year={2024},
  publisher={ACM New York, NY, USA}
}

@article{xia2023conversational,
  title={Conversational automated program repair},
  author={Xia, Chunqiu Steven and Zhang, Lingming},
  journal={arXiv preprint arXiv:2301.13246},
  year={2023}
}

@inproceedings{xia2024automated,
  title={Automated program repair via conversation: Fixing 162 out of 337 bugs for \$0.42 each using chatgpt},
  author={Xia, Chunqiu Steven and Zhang, Lingming},
  booktitle={Proceedings of the 33rd ACM SIGSOFT International Symposium on Software Testing and Analysis},
  pages={819--831},
  year={2024}
}

@article{chen2021codex,
  title={Evaluating large language models trained on code},
      author={Mark Chen and Jerry Tworek and Heewoo Jun and Qiming Yuan and Henrique Ponde de Oliveira Pinto and Jared Kaplan and Harri Edwards and Yuri Burda and Nicholas Joseph and Greg Brockman and Alex Ray and Raul Puri and Gretchen Krueger and Michael Petrov and Heidy Khlaaf and Girish Sastry and Pamela Mishkin and Brooke Chan and Scott Gray and Nick Ryder and Mikhail Pavlov and Alethea Power and Lukasz Kaiser and Mohammad Bavarian and Clemens Winter and Philippe Tillet and Felipe Petroski Such and Dave Cummings and Matthias Plappert and Fotios Chantzis and Elizabeth Barnes and Ariel Herbert-Voss and William Hebgen Guss and Alex Nichol and Alex Paino and Nikolas Tezak and Jie Tang and Igor Babuschkin and Suchir Balaji and Shantanu Jain and William Saunders and Christopher Hesse and Andrew N. Carr and Jan Leike and Josh Achiam and Vedant Misra and Evan Morikawa and Alec Radford and Matthew Knight and Miles Brundage and Mira Murati and Katie Mayer and Peter Welinder and Bob McGrew and Dario Amodei and Sam McCandlish and Ilya Sutskever and Wojciech Zaremba},
  journal={arXiv preprint arXiv:2107.03374},
  year={2021}
}

@article{zhang2025swe,
  title={SWE-bench Goes Live!},
  author={Linghao Zhang and Shilin He and Chaoyun Zhang and Yu Kang and Bowen Li and Chengxing Xie and Junhao Wang and Maoquan Wang and Yufan Huang and Shengyu Fu and Elsie Nallipogu and Qingwei Lin and Yingnong Dang and Saravan Rajmohan and Dongmei Zhang},
  journal={arXiv preprint arXiv:2505.23419},
  year={2025}
}

@article{liu2024marscode,
  title={Marscode agent: Ai-native automated bug fixing},
  author={Liu, Yizhou and Gao, Pengfei and Wang, Xinchen and Liu, Jie and Shi, Yexuan and Zhang, Zhao and Peng, Chao},
  journal={arXiv preprint arXiv:2409.00899},
  year={2024}
}

@article{kong2025contrastrepair,
  title={Contrastrepair: Enhancing conversation-based automated program repair via contrastive test case pairs},
  author={Kong, Jiaolong and Xie, Xiaofei and Cheng, Mingfei and Liu, Shangqing and Du, Xiaoning and Guo, Qi},
  journal={ACM Transactions on Software Engineering and Methodology},
  volume={34},
  number={8},
  pages={1--31},
  year={2025}
}

@article{chen2023teaching,
  title={Teaching large language models to self-debug},
  author={Chen, Xinyun and Lin, Maxwell and Sch{\"a}rli, Nathanael and Zhou, Denny},
  journal={arXiv preprint arXiv:2304.05128},
  year={2023}
}

@misc{luodeepswe,
  title={Deepswe: Training a fully open-sourced, state-of-the-art coding agent by scaling rl},
  author={Luo, Michael and Jain, Naman and Singh, Jaskirat and Tan, Sijun and Patel, Ameen and Wu, Qingyang and Ariyak, Alpay and Cai, Colin and Venkat, Shang Zhu Tarun and Athiwaratkun, Ben and others}
}

@misc{swe15,
    title = {{SWE-1.5}},
author={Cognition},
    key = {aaa},
    note = "\url{https://cognition.ai/blog/swe-1-5}"
}

@misc{deepseekv31,
    title = {{DeepSeek V3.1}},
author={DeepSeek AI},
    key = {aaa},
    note = "\url{https://api-docs.deepseek.com/news/news250821}"
}

@article{chen2025minimax,
  title={MiniMax-M1: Scaling Test-Time Compute Efficiently with Lightning Attention},
  author={Chen, Aili and Li, Aonian and Gong, Bangwei and Jiang, Binyang and Fei, Bo and Yang, Bo and Shan, Boji and Yu, Changqing and Wang, Chao and Zhu, Cheng and others},
  journal={arXiv preprint arXiv:2506.13585},
  year={2025}
}

@misc{sbm,
    title = {SWE-bench Multilingual},
author="Kabir Khandpur and Kilian Lieret and Carlos E. Jimenez and Ofir Press and John Yang",
year=2025,
    key = {aaa},
    note = "\url{https://www.swebench.com/multilingual.html}"
}

@misc{moatless,
    title = {Moatless},
author="Albert Örwall",
year=2024,
    key = {aaa},
    note = "\url{https://github.com/aorwall/moatless-tools}"
}

@article{yang2024swe,
  title={Swe-agent: Agent-computer interfaces enable automated software engineering},
  author={Yang, John and Jimenez, Carlos E and Wettig, Alexander and Lieret, Kilian and Yao, Shunyu and Narasimhan, Karthik and Press, Ofir},
  journal={Advances in Neural Information Processing Systems},
  volume={37},
  pages={50528--50652},
  year={2024}
}

@article{wang2024openhands,
  title={Openhands: An open platform for ai software developers as generalist agents},
  author={Xingyao Wang and Boxuan Li and Yufan Song and Frank F. Xu and Xiangru Tang and Mingchen Zhuge and Jiayi Pan and Yueqi Song and Bowen Li and Jaskirat Singh and Hoang H. Tran and Fuqiang Li and Ren Ma and Mingzhang Zheng and Bill Qian and Yanjun Shao and Niklas Muennighoff and Yizhe Zhang and Binyuan Hui and Junyang Lin and Robert Brennan and Hao Peng and Heng Ji and Graham Neubig},
  journal={arXiv preprint arXiv:2407.16741},
  year={2024}
}

@inproceedings{zhang2024autocoderover,
  title={Autocoderover: Autonomous program improvement},
  author={Zhang, Yuntong and Ruan, Haifeng and Fan, Zhiyu and Roychoudhury, Abhik},
  booktitle={Proceedings of the 33rd ACM SIGSOFT International Symposium on Software Testing and Analysis},
  pages={1592--1604},
  year={2024}
}

@article{team2025kimi,
  title={Kimi k1.5: Scaling reinforcement learning with llms},
  author={Kimi Team},
  journal={arXiv preprint arXiv:2501.12599},
  year={2025}
}

@article{jimenez2023swe,
  title={Swe-bench: Can language models resolve real-world github issues?},
  author={Jimenez, Carlos E and Yang, John and Wettig, Alexander and Yao, Shunyu and Pei, Kexin and Press, Ofir and Narasimhan, Karthik},
  journal={arXiv preprint arXiv:2310.06770},
  year={2023}
}

@article{rashid2025swe,
  title={SWE-PolyBench: A multi-language benchmark for repository level evaluation of coding agents},
  author={Muhammad Shihab Rashid and Christian Bock and Yuan Zhuang and Alexander Buchholz and Tim Esler and Simon Valentin and Luca Franceschi and Martin Wistuba and Prabhu Teja Sivaprasad and Woo Jung Kim and Anoop Deoras and Giovanni Zappella and Laurent Callot},
  journal={arXiv preprint arXiv:2504.08703},
  year={2025}
}

@article{zan2025multi,
  title={Multi-swe-bench: A multilingual benchmark for issue resolving},
  author={Daoguang Zan and Zhirong Huang and Wei Liu and Hanwu Chen and Linhao Zhang and Shulin Xin and Lu Chen and Qi Liu and Xiaojian Zhong and Aoyan Li and Siyao Liu and Yongsheng Xiao and Liangqiang Chen and Yuyu Zhang and Jing Su and Tianyu Liu and Rui Long and Kai Shen and Liang Xiang},
  journal={arXiv preprint arXiv:2504.02605},
  year={2025}
}

@misc{sbv,
  title = "{SWE-bench Verified}",
  author = "OpenAI",
  key = "aaa",
year=2025,
  note = "\url{https://openai.com/index/introducing-swe-bench-verified/}"
}

@misc{sonnet4.5,
  title = "{Claude Sonnet 4.5}",
  author = "Anthropic",
  key = "aaa",
year=2025,
  note = "\url{https://www.anthropic.com/news/claude-sonnet-4-5}"
}

@article{deng2025swe,
  title={SWE-Bench Pro: Can AI Agents Solve Long-Horizon Software Engineering Tasks?},
  author={Scale AI},
  journal={arXiv preprint arXiv:2509.16941},
  year={2025}
}

@misc{gpt51,
  title = "{GPT-5.1}",
  key = "aaa",
  note = "\url{https://openai.com/index/gpt-5-1-for-developers/}"
}

@misc{devin,
  title = "{Devin AI}",
  key = "aaa",
  author="Cognition",
year=2024,
  note = "\url{https://cognition.ai/blog/introducing-devin}"
}

@article{austin2021program,
    title={Program synthesis with large language models},
    author={Austin, Jacob and Odena, Augustus and Nye, Maxwell and Bosma, Maarten and Michalewski, Henryk and Dohan, David and Jiang, Ellen and Cai, Carrie and Terry, Michael and Le, Quoc and Sutton, Charles},
    journal={arXiv preprint arXiv:2108.07732},
    year={2021}
}

@misc{kimik2thinking,
    title = {kimi-k2-thinking},
  key = "aaa",
    note = "\url{https://moonshotai.github.io/Kimi-K2/thinking.html}"
}

@misc{wang2025huxley,
      title={Huxley-G\"odel Machine: Human-Level Coding Agent Development by an Approximation of the Optimal Self-Improving Machine}, 
      author={Wenyi Wang and Piotr Piękos and Li Nanbo and Firas Laakom and Yimeng Chen and Mateusz Ostaszewski and Mingchen Zhuge and Jürgen Schmidhuber},
      year={2025},
      eprint={2510.21614},
      archivePrefix={arXiv},
      primaryClass={cs.AI},
      url={https://arxiv.org/abs/2510.21614}, 
}

@article{badertdinov2025swe,
  title={SWE-rebench: An Automated Pipeline for Task Collection and Decontaminated Evaluation of Software Engineering Agents},
  author={Badertdinov, Ibragim and Golubev, Alexander and Nekrashevich, Maksim and Shevtsov, Anton and Karasik, Simon and Andriushchenko, Andrei and Trofimova, Maria and Litvintseva, Daria and Yangel, Boris},
  journal={arXiv preprint arXiv:2505.20411},
  year={2025}
}

@inproceedings{deng2023large,
  title={Large language models are zero-shot fuzzers: Fuzzing deep-learning libraries via large language models},
  author={Deng, Yinlin and Xia, Chunqiu Steven and Peng, Haoran and Yang, Chenyuan and Zhang, Lingming},
  booktitle={Proceedings of the 32nd ACM SIGSOFT international symposium on software testing and analysis},
  pages={423--435},
  year={2023}
}

@inproceedings{xia2022less,
  title={Less training, more repairing please: revisiting automated program repair via zero-shot learning},
  author={Xia, Chunqiu Steven and Zhang, Lingming},
  booktitle={Proceedings of the 30th ACM Joint European Software Engineering Conference and Symposium on the Foundations of Software Engineering},
  pages={959--971},
  year={2022}
}

@article{liu2024large,
  title={Large language model-based agents for software engineering: A survey},
  author={Liu, Junwei and Wang, Kaixin and Chen, Yixuan and Peng, Xin and Chen, Zhenpeng and Zhang, Lingming and Lou, Yiling},
  journal={arXiv preprint arXiv:2409.02977},
  year={2024}
}

@inproceedings{yang2024sweagent,
  title={{SWE}-agent: Agent-Computer Interfaces Enable Automated Software Engineering},
  author={John Yang and Carlos E Jimenez and Alexander Wettig and Kilian Lieret and Shunyu Yao and Karthik R Narasimhan and Ofir Press},
  booktitle={The Thirty-eighth Annual Conference on Neural Information Processing Systems},
  year={2024},
  url={https://arxiv.org/abs/2405.15793}
}

@inproceedings{
    robeyns2025sica,
    title={{SICA} A Self-Improving Coding Agent},
    author={Maxime Robeyns and Martin Szummer and Laurence Aitchison},
    booktitle={ICLR 2025 Workshop on Scaling Self-Improving Foundation Models},
    year={2025},
    url={https://openreview.net/forum?id=rShJCyLsOr}
}

@article{zhang2025darwin,
  title={Darwin Godel Machine: Open-Ended Evolution of Self-Improving Agents},
  author={Zhang, Jenny and Hu, Shengran and Lu, Cong and Lange, Robert and Clune, Jeff},
  journal={arXiv preprint arXiv:2505.22954},
  year={2025}
}

@article{li2022competition,
  title={Competition-level code generation with alphacode},
  author={Yujia Li and David Choi and Junyoung Chung and Nate Kushman and Julian Schrittwieser and Rémi Leblond and Tom Eccles and James Keeling and Felix Gimeno and Agustin Dal Lago and Thomas Hubert and Peter Choy and Cyprien de Masson d'Autume and Igor Babuschkin and Xinyun Chen and Po-Sen Huang and Johannes Welbl and Sven Gowal and Alexey Cherepanov and James Molloy and Daniel J. Mankowitz and Esme Sutherland Robson and Pushmeet Kohli and Nando de Freitas and Koray Kavukcuoglu and Oriol Vinyals},
  journal={arXiv preprint arXiv:2203.07814},
  year={2022}
}

@article{hinton2002stochastic,
  title={Stochastic neighbor embedding},
  author={Hinton, Geoffrey E and Roweis, Sam},
  journal={Advances in neural information processing systems},
  volume={15},
  year={2002}
}

@misc{skills,
    title = {{Introducing Agent Skills}},
author={Anthropic},
    key = {aaa},
    note = "\url{https://www.claude.com/blog/skills}"
}

@inproceedings{lee2025sec,
  author    = {Hwiwon Lee and Ziqi Zhang and Hanxiao Lu and Lingming Zhang},
  booktitle = {The Thirty-ninth Annual Conference on Neural Information Processing Systems},
  title     = {{SEC-bench: Automated Benchmarking of LLM Agents on Real-World Software Security Tasks}},
  url       = {https://openreview.net/forum?id=QQhQIqons0},
  year      = {2025}
}

@article{zhao2024commit0,
  title={Commit0: Library generation from scratch},
  author={Zhao, Wenting and Jiang, Nan and Lee, Celine and Chiu, Justin T and Cardie, Claire and Gall{\'e}, Matthias and Rush, Alexander M},
  journal={arXiv preprint arXiv:2412.01769},
  year={2024}
}

@inproceedings{llmastool,
 author = {Cai, Tianle and Wang, Xuezhi and Ma, Tengyu and Chen, Xinyun and Zhou, Denny},
 booktitle = {International Conference on Representation Learning},
 editor = {B. Kim and Y. Yue and S. Chaudhuri and K. Fragkiadaki and M. Khan and Y. Sun},
 pages = {54067--54089},
 title = {Large Language Models as Tool Makers},
 url = {https://proceedings.iclr.cc/paper_files/paper/2024/file/ed91353f700d113e5d848c7e04a858b0-Paper-Conference.pdf},
 volume = {2024},
 year = {2024}
}

@article{
wang2024voyager,
title={Voyager: An Open-Ended Embodied Agent with Large Language Models},
author={Guanzhi Wang and Yuqi Xie and Yunfan Jiang and Ajay Mandlekar and Chaowei Xiao and Yuke Zhu and Linxi Fan and Anima Anandkumar},
journal={Transactions on Machine Learning Research},
issn={2835-8856},
year={2024},
url={https://openreview.net/forum?id=ehfRiF0R3a},
note={}
}

@misc{qiu2025alitageneralistagentenabling,
      title={Alita: Generalist Agent Enabling Scalable Agentic Reasoning with Minimal Predefinition and Maximal Self-Evolution}, 
      author={Jiahao Qiu and Xuan Qi and Tongcheng Zhang and Xinzhe Juan and Jiacheng Guo and Yifu Lu and Yimin Wang and Zixin Yao and Qihan Ren and Xun Jiang and Xing Zhou and Dongrui Liu and Ling Yang and Yue Wu and Kaixuan Huang and Shilong Liu and Hongru Wang and Mengdi Wang},
      year={2025},
      eprint={2505.20286},
      archivePrefix={arXiv},
      primaryClass={cs.AI},
      url={https://arxiv.org/abs/2505.20286}, 
}

@misc{qian2024creatortoolcreationdisentangling,
      title={CREATOR: Tool Creation for Disentangling Abstract and Concrete Reasoning of Large Language Models}, 
      author={Cheng Qian and Chi Han and Yi R. Fung and Yujia Qin and Zhiyuan Liu and Heng Ji},
      year={2024},
      eprint={2305.14318},
      archivePrefix={arXiv},
      primaryClass={cs.CL},
      url={https://arxiv.org/abs/2305.14318}, 
}

@misc{wang2024troveinducingverifiableefficient,
      title={TroVE: Inducing Verifiable and Efficient Toolboxes for Solving Programmatic Tasks}, 
      author={Zhiruo Wang and Daniel Fried and Graham Neubig},
      year={2024},
      eprint={2401.12869},
      archivePrefix={arXiv},
      primaryClass={cs.AI},
      url={https://arxiv.org/abs/2401.12869}, 
}

\clearpage
\newpage
\appendix

\section{Additional experimental settings}
\label{appendix:additional_experimental_settings}

We now describe additional experimental settings used by \tech for evaluation.
As discussed in Section~\ref{sec:setup}, we implement \tech on top of the \minisweagent framework.
We use the same default hyperparameter of \minisweagent at the time of writing (maximum step limit of 250 and maximum cost of \$3 per issue).

For \geminithreepro{}, we use temperature of 1 since it is recommended by the developers\footnote{\url{https://ai.google.dev/gemini-api/docs/gemini-3}}.
For any Anthropic models in our experiments, we follow \minisweagent and use a temperature of 0.0.
For any OpenAI models in our experiments (e.g., \gptfive, \gptfivemini, \gptfivenano), since temperature 0.0 sampling is not supported, we use temperature of 1.
To the best of our knowledge, this is also the case used in \minisweagent when they use OpenAI models.

\section{Additional tool analysis}
\label{appendix:tool_analysis}

\begin{figure*}[h]
    \centering
    \begin{subfigure}[b]{0.32\textwidth}
    \centering
    \includegraphics[width=\textwidth]{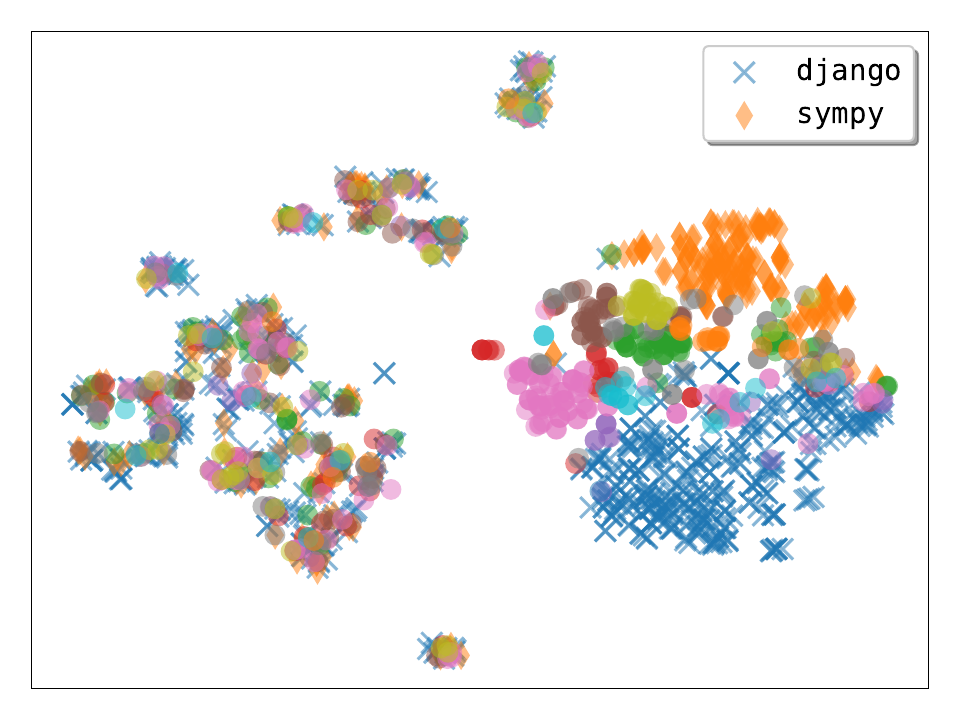}
    \caption{Tools in \swebench Verified}
    \label{fig:tool_verified_repo}
\end{subfigure}
\begin{subfigure}[b]{0.32\textwidth}
    \centering
    \includegraphics[width=\textwidth]{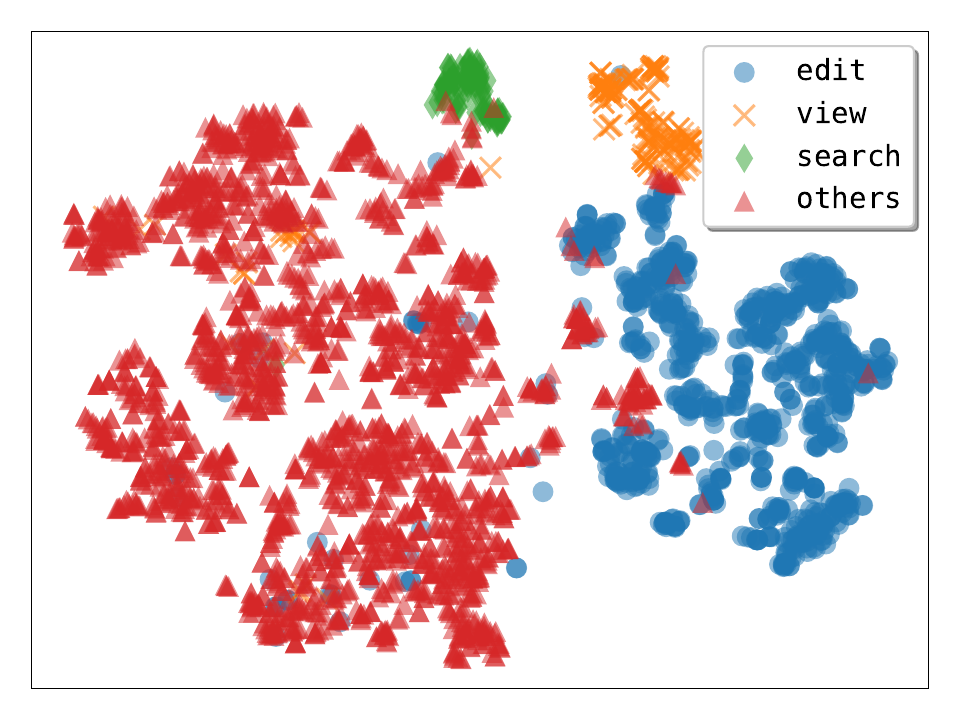}
    \caption{Tools in \swebenchpro{}}
    \label{fig:tool_pro_tool_name}
\end{subfigure}
\caption{
Additional 2 dimensional t-SNE visualization of tools generated by \claudesonnetfourfive{} on \swebench Verified and \swebenchpro.
We label and display the embedding based on the repository name (Figure~\ref{fig:tool_verified_repo}) and tool name (Figure~\ref{fig:tool_pro_tool_name}).
Note that for Figure~\ref{fig:tool_verified_repo}, we only label two repositories in the legend due to space considerations. 
The two repositories were chosen as they have representative distinct clusters.
}
\end{figure*}

We perform additional tool analysis similar to the ones done in Section~\ref{sec:tool_analysis}.
We added the missing results including the visualization based on repository used for \swebench Verified and tool name for \swebenchpro.
Note that we did not include any results based on programming language for \swebench Verified since they are all repositories in Python.
We observe similar patterns to the main evaluation.

To perform this analysis, we use \CodeIn{sklearn} implementation of PCA with \CodeIn{n\_components} of 50. 
We then use the t-SNE implementation from \CodeIn{sklearn} with \CodeIn{max\_iter} of 1000.

\section{Ablation problems}

\subsection{\swebench Verified ablation subset}
\label{appendix:ablation_dataset}

Here we include the 50 randomly selected problems in \swebench Verified used in our ablation evaluations.

\begin{multicols}{2}
\begin{itemize}
\item \texttt{sympy\_\_sympy-12489}
\item \texttt{django\_\_django-16502}
\item \texttt{scikit-learn\_\_scikit-learn-11310}
\item \texttt{django\_\_django-13590}
\item \texttt{scikit-learn\_\_scikit-learn-11578}
\item \texttt{matplotlib\_\_matplotlib-21568}
\item \texttt{django\_\_django-11276}
\item \texttt{django\_\_django-11119}
\item \texttt{sympy\_\_sympy-13031}
\item \texttt{sphinx-doc\_\_sphinx-8459}
\item \texttt{django\_\_django-14170}
\item \texttt{django\_\_django-11066}
\item \texttt{sphinx-doc\_\_sphinx-9658}
\item \texttt{django\_\_django-14631}
\item \texttt{django\_\_django-11749}
\item \texttt{scikit-learn\_\_scikit-learn-13135}
\item \texttt{astropy\_\_astropy-14369}
\item \texttt{matplotlib\_\_matplotlib-24177}
\item \texttt{django\_\_django-15252}
\item \texttt{pydata\_\_xarray-7233}
\item \texttt{django\_\_django-11815}
\item \texttt{sympy\_\_sympy-18199}
\item \texttt{django\_\_django-15467}
\item \texttt{psf\_\_requests-5414}
\item \texttt{django\_\_django-17084}
\item \texttt{scikit-learn\_\_scikit-learn-15100}
\item \texttt{django\_\_django-13925}
\item \texttt{django\_\_django-11163}
\item \texttt{django\_\_django-16595}
\item \texttt{django\_\_django-16938}
\item \texttt{sympy\_\_sympy-13877}
\item \texttt{django\_\_django-16612}
\item \texttt{django\_\_django-15629}
\item \texttt{pydata\_\_xarray-3677}
\item \texttt{django\_\_django-15503}
\item \texttt{psf\_\_requests-1142}
\item \texttt{mwaskom\_\_seaborn-3069}
\item \texttt{django\_\_django-15127}
\item \texttt{astropy\_\_astropy-13033}
\item \texttt{django\_\_django-16429}
\item \texttt{django\_\_django-16256}
\item \texttt{django\_\_django-11133}
\item \texttt{django\_\_django-13741}
\item \texttt{matplotlib\_\_matplotlib-23476}
\item \texttt{scikit-learn\_\_scikit-learn-12973}
\item \texttt{scikit-learn\_\_scikit-learn-13779}
\item \texttt{sympy\_\_sympy-23413}
\item \texttt{pytest-dev\_\_pytest-7521}
\item \texttt{sympy\_\_sympy-22456}
\item \texttt{django\_\_django-16082}
\end{itemize}
\end{multicols}

\subsection{\swebenchmultilingual ablation subset}
\label{appendix:multilingal_subset}

We further evaluate on 50 randomly selected subset of problems in \swebenchmultilingual. 
Below is the list of \CodeIn{instance\_ids} of the problems in our subset.

\begin{multicols}{2}
\begin{itemize}
\item \texttt{apache\_\_druid-13704}
\item \texttt{apache\_\_druid-14136}
\item \texttt{apache\_\_druid-15402}
\item \texttt{apache\_\_lucene-13494}
\item \texttt{astral-sh\_\_ruff-15543}
\item \texttt{axios\_\_axios-6539}
\item \texttt{babel\_\_babel-15445}
\item \texttt{burntsushi\_\_ripgrep-2576}
\item \texttt{caddyserver\_\_caddy-5404}
\item \texttt{caddyserver\_\_caddy-5995}
\item \texttt{caddyserver\_\_caddy-6051}
\item \texttt{facebook\_\_docusaurus-9183}
\item \texttt{faker-ruby\_\_faker-2970}
\item \texttt{fastlane\_\_fastlane-20642}
\item \texttt{fluent\_\_fluentd-3640}
\item \texttt{fmtlib\_\_fmt-3750}
\item \texttt{javaparser\_\_javaparser-4538}
\item \texttt{javaparser\_\_javaparser-4561}
\item \texttt{jqlang\_\_jq-2919}
\item \texttt{laravel\_\_framework-48636}
\item \texttt{laravel\_\_framework-52451}
\item \texttt{mrdoob\_\_three.js-27395}
\item \texttt{nushell\_\_nushell-12901}
\item \texttt{phpoffice\_\_phpspreadsheet-3659}
\item \texttt{preactjs\_\_preact-2757}
\item \texttt{preactjs\_\_preact-2927}
\item \texttt{preactjs\_\_preact-3562}
\item \texttt{preactjs\_\_preact-3739}
\item \texttt{projectlombok\_\_lombok-3326}
\item \texttt{reactivex\_\_rxjava-7597}
\item \texttt{redis\_\_redis-10068}
\item \texttt{redis\_\_redis-10095}
\item \texttt{rubocop\_\_rubocop-13503}
\item \texttt{rubocop\_\_rubocop-13560}
\item \texttt{rubocop\_\_rubocop-13627}
\item \texttt{rubocop\_\_rubocop-13680}
\item \texttt{rubocop\_\_rubocop-13705}
\item \texttt{sharkdp\_\_bat-2650}
\item \texttt{sharkdp\_\_bat-562}
\item \texttt{tokio-rs\_\_axum-1730}
\item \texttt{tokio-rs\_\_axum-2096}
\item \texttt{tokio-rs\_\_axum-682}
\item \texttt{tokio-rs\_\_tokio-6603}
\item \texttt{tokio-rs\_\_tokio-6752}
\item \texttt{tokio-rs\_\_tokio-6838}
\item \texttt{uutils\_\_coreutils-6731}
\item \texttt{valkey-io\_\_valkey-1842}
\item \texttt{valkey-io\_\_valkey-790}
\item \texttt{vuejs\_\_core-11739}
\item \texttt{vuejs\_\_core-11870}
\end{itemize}
\end{multicols}

\section{Prompts Used}
\label{appendix:initial_prompt}

Here we provide the detailed prompts used by \tech. 
Note that the majority of the prompt is constructed by \minisweagent and we only make minimal changes to enable on-the-fly tool creation.
Furthermore, we perform a slight modification: change the home directory indication of \CodeIn{testbed} to \CodeIn{app} in the initial prompt for all \swebenchpro evaluations.
This is done since the \swebenchpro docker environments use \CodeIn{app} as the main source directory containing the repository code, different from \swebench environments.
Apart from that, we use the same prompt for all our evaluations.

\subsection{Initial prompt}

\begin{promptbox}{Initial prompt for \tech{}}
    \begin{Verbatim}[breaklines=true]
<pr_description>
Consider the following PR description:
{{task}}
</pr_description>

<instructions>
# Task Instructions

## Overview
You're a software engineer interacting continuously with a computer by submitting commands.
You'll be helping implement necessary changes to meet requirements in the PR description.
Your task is specifically to make changes to non-test files in the current directory in order to fix the issue described in the PR description in a way that is general and consistent with the codebase.

IMPORTANT: This is an interactive process where you will think and issue ONE command, see its result, then think and issue your next command.

For each response:
1. Include a THOUGHT section explaining your reasoning and what you're trying to accomplish
2. Provide exactly ONE bash command to execute

## Important Boundaries
- MODIFY: Regular source code files in /testbed (this is the working directory for all your subsequent commands)
- DO NOT MODIFY: Tests, configuration files (pyproject.toml, setup.cfg, etc.)

## Recommended Workflow
1. Analyze the codebase by finding and reading relevant files
2. Create a script to reproduce the issue
3. Edit the source code to resolve the issue
4. Verify your fix works by running your script again
5. Test edge cases to ensure your fix is robust

## Command Execution Rules
You are operating in an environment where
1. You write a single command
2. The system executes that command in a subshell
3. You see the result
4. You write your next command

Each response should include:
1. A **THOUGHT** section where you explain your reasoning and plan
2. A single bash code block with your command

Format your responses like this:

<format_example>
THOUGHT: Here I explain my reasoning process, analysis of the current situation,
and what I'm trying to accomplish with the command below.

```bash
your_command_here
```
</format_example>

Commands must be specified in a single bash code block:

```bash
your_command_here
```

**CRITICAL REQUIREMENTS:**
- Your response SHOULD include a THOUGHT section explaining your reasoning
- Your response MUST include EXACTLY ONE bash code block
- This bash block MUST contain EXACTLY ONE command (or a set of commands connected with && or ||)
- If you include zero or multiple bash blocks, or no command at all, YOUR RESPONSE WILL FAIL
- Do NOT try to run multiple independent commands in separate blocks in one response
- Directory or environment variable changes are not persistent. Every action is executed in a new subshell.
- However, you can prefix any action with `MY_ENV_VAR=MY_VALUE cd /path/to/working/dir && ...` or write/load environment variables from files

Example of a CORRECT response:
<example_response>
THOUGHT: I need to understand the structure of the repository first. Let me check what files are in the current directory to get a better understanding of the codebase.

```bash
ls -la
```
</example_response>

Example of an INCORRECT response:
<example_response>
THOUGHT: I need to examine the codebase and then look at a specific file. I'll run multiple commands to do this.

```bash
ls -la
```

Now I'll read the file:

```bash
cat file.txt
```
</example_response>

If you need to run multiple commands, either:
1. Combine them in one block using && or ||
```bash
command1 && command2 || echo "Error occurred"
```

2. Wait for the first command to complete, see its output, then issue the next command in your following response.

## Environment Details
- You have a full Linux shell environment
- Always use non-interactive flags (-y, -f) for commands
- Avoid interactive tools like vi, nano, or any that require user input
- If a command isn't available, you can install it

## Useful Command Examples

### Create a new file:
```bash
cat <<'EOF' > newfile.py
import numpy as np
hello = "world"
print(hello)
EOF
```

### View file content:
```bash
# View specific lines with numbers
nl -ba filename.py | sed -n '10,20p'
```

**IMPORTANT TOOL CREATION INSTRUCTIONS**
## Creating your own tools 
- You can also create your own tools in Python to help with your workflow
- Compared to basic bash commands, the tools you create should be able to better aid your workflow in solving the task
- Ensure each tool you create is in Python, contains informative outputs or error messages, and can be ran from the command line
- You should at least create a simple edit tool that can help you effectively edit arbitrary files instead of using bash commands
- The tools you create can be for any purpose, it does not need to be general, instead think about how it can help you specifically with the current task at hand

### Example of creating a custom tool:
<example_response>
THOUGHT: I noticed that in order to solve the issue I need to ... therefore I should create a custom tool to help me ...

```bash
cat <<'EOF' > /path/to/tool_name.py
#!/usr/bin/env python3
import sys
# Import other packages if needed

def main():
    # Your tool logic here
    ...

if __name__ == "__main__":
    main()
EOF
```
</example_response>

### Example of using the tool you created:
<example_response>
THOUGHT: Let me use the custom tool I created to help me with ...

```bash
python /path/to/tool_name.py <<EOF
your_input_here
EOF
```
</example_response>

## Submission
When you've completed your work (reading, editing, testing), and cannot make further progress
issue exactly the following command:

```bash
echo COMPLETE_TASK_AND_SUBMIT_FINAL_OUTPUT && git add -A && git diff --cached
```

This command will submit your work.
You cannot continue working (reading, editing, testing) in any way on this task after submitting.
</instructions>
\end{Verbatim}
\end{promptbox}
\captionsetup{type=figure}
\captionof{figure}{
The initial prompt for \tech, modified from the base \minisweagent prompt.
}
\label{fig:initial_prompt}

\subsection{Feedback message}

\begin{promptbox}{Feedback message used by \tech{}}
    \begin{Verbatim}[breaklines=true]
<returncode>{{output.returncode}}</returncode>
{% if output.output | length < 10000 -%}
<output>
{{ output.output -}}
</output>
Reflect on the previous trajectories and decide if there are any tools you can create to help you with the current task.
Note that just because you can use basic bash commands doesn't mean you should not create any tools that can still be helpful.
{%- else -%}
<warning>
The output of your last command was too long.
Please try a different command that produces less output.
If you're looking at a file you can try use head, tail, sed or create a tool to view a smaller number of lines selectively.
If you're using grep or find and it produced too much output, you can use a more selective search pattern.
</warning>
{%- set elided_chars = output.output | length - 10000 -%}
<output_head>
{{ output.output[:5000] }}
</output_head>
<elided_chars>
{{ elided_chars }} characters elided
</elided_chars>
<output_tail>
{{ output.output[-5000:] }}
</output_tail>
{%- endif -%}
\end{Verbatim}
\end{promptbox}
\captionsetup{type=figure}
\captionof{figure}{
The feedback message used by \tech after each agent step.
}
\label{fig:feedback_prompt}

\end{document}